\documentclass[prd,twocolumn,showpacs,floatfix,superscriptaddress,nofootinbib]{revtex4}
\usepackage[utf8]{inputenc}
\usepackage{graphicx}
\usepackage{epsfig}
\usepackage{bm}
\usepackage{amsfonts}
\usepackage[T1]{fontenc}
\usepackage{amssymb}
\usepackage{float}
\usepackage{amsmath}
\usepackage{dcolumn}
\usepackage{wasysym}
\usepackage{cancel}
\usepackage[colorlinks]{hyperref}
\usepackage[usenames,dvipsnames]{color}
\hypersetup{
     breaklinks=true,
    pdfstartview={FitH},    
    colorlinks=true,       
    linkcolor=blue,          
    citecolor=red,        
    filecolor=magenta,      
    urlcolor=blue,           
    anchorcolor=green,      
    linktocpage=true
}

\def\doi{http://doi.org}

\def\be{\begin{equation*}}
\def\ee{\end{equation*}}

\begin{document}

\title{Massive 
Particle Motion Around  Horndeski Black Holes}
\author{D. A. Carvajal}
\email{diego.carvajala@usach.cl} \affiliation{Instituto de F\'{i}sica, 
 Pontificia Universidad Cat\'{o}lica de Chile, Avenida Vicuña Mackenna 4860, Santiago, Chile.}
 \affiliation{Departamento de Matemática y Ciencia de la Computación, Universidad de Santiago de Chile, Las Sophoras 173, Santiago, Chile.}
\author{P. A. Gonz\'{a}lez}
\email{pablo.gonzalez@udp.cl} \affiliation{Facultad de
Ingenier\'{i}a y Ciencias, Universidad Diego Portales, Avenida Ej\'{e}rcito
Libertador 441, Casilla 298-V, Santiago, Chile.}
\author{Marco Olivares}
\email{marco.olivaresr@mail.udp.cl}
\affiliation{Facultad de
Ingenier\'{i}a y Ciencias, Universidad Diego Portales, Avenida Ej\'{e}rcito
Libertador 441, Casilla 298-V, Santiago, Chile.}
\author{Eleftherios Papantonopoulos}
\email{lpapa@central.ntua.gr}
\affiliation{Physics Division, School of Applied Mathematical and Physical Sciences, National Technical University of Athens, 15780 Zografou Campus, Athens, Greece.}
\author{Yerko V\'{a}squez}
\email{yvasquez@userena.cl}
\affiliation{Departamento de F\'{\i}sica, Facultad de Ciencias, Universidad de La Serena,\\
Avenida Cisternas 1200, La Serena, Chile.}
\date{\today}



\begin{abstract}

The time-like structure of the four-dimensional 
asymptotically flat Horndeski black holes is studied in detail. Focusing on the motion of massive neutral test particles, we construct the corresponding effective potential and classify the admissible types of orbits. The equations of motion are solved analytically, yielding trajectories expressed in terms of Weierstra\ss \, elliptic functions and elementary functions. As an application, we compute the perihelion precession as a classical test of gravity within the Solar System and use it to place observational constraints on the coupling parameter between the scalar field and gravity.

\end{abstract}

\maketitle


\tableofcontents




\section{Introduction}

In General Relativity (GR) black holes are celestial compact objects and they have been extensively studied. The recent observational results on the
collision of two black holes or neutron stars, and the generation of gravitational waves (GWs) \cite{Abbott:2016blz}-\cite{TheLIGOScientific:2017qsa}, gave us a new understanding of their formation and evaporation in extreme gravity conditions. The observation of a shadow of the M 87 black hole from the Event Horizon Telescope
\cite{EventHorizonTelescope:2019dse} confirmed our confidence that GR is a very successful viable theory. We also note that the recent observational results on dark matter and on dark energy gave us new information about the generation and expansion of the Universe.  However, to have a viable theory of GR on short and large distances explaining these results on cosmological grounds, a generalization of GR is required \cite{Joyce:2014kja,Nojiri:2006ri,Clifton:2011jh,MG3}. 

Theories that modify GR were introduced to give us important information on the structure and properties of compact objects predicted by these theories and also to give observational signatures, which are consistent with the recent observations.
Viable modifications of GR which have been intensely studied are resulting from the presence of high curvature terms and from scalar fields coupled to gravity backreacting to the background metric, dressing the black hole solutions with hair, and are known as scalar-tensor theories \cite{Fujii:2003pa}. A crucial question in the scalar-tensor theories is related to their structure, the behavior of black hole solutions, and their stability \cite{Martinez:1996gn}-\cite{Babichev:2013cya}.

One of the most general scalar-tensor theory with second derivative equations of motion is the Horndeski theory \cite{Horndeski:1974wa}. Soon it was realized that Horndeski theory is not the only scalar-tensor theory and a new class of gravitational scalar-tensor theories was proposed \cite{Gleyzes:2014dya,Gleyzes:2014qga} which are generalizations of Horndeski theories. Higher derivative equations of motion generically lead to ghost instabilities \cite{Ostrogradsky:1850fid}. However,  these theories can still be physically
sensible. It was shown that there exist theories with higher
derivative equations of motion that are degenerate and still have the canonical number of
propagating degrees of freedom, thus evading ghosts \cite{Gleyzes:2014dya,Gleyzes:2014qga,Zumalacarregui:2013pma,Deffayet:2015qwa,Crisostomi:2016tcp,Langlois:2015skt,BenAchour:2016fzp}. 

The Horndeski theory has been analyzed across both short and cosmological distances. At shorter scales, black holes and wormholes arise from a gravitational
action with a real or phantom scalar field, respectively, non-minimally coupled to the Einstein tensor \cite{Kolyvaris:2011fk,Rinaldi:2012vy,Kolyvaris:2013zfa,Babichev:2013cya,Charmousis:2014zaa,Korolev:2014hwa}. In these exact solutions, the 'gravitational' scalar and its coupling strength appear in
the metric components as a primary charge, which at large distances plays the role of an effective cosmological constant.  On cosmological scales, the derivative coupling in Horndeski theories acts as a friction term during the inflationary period of the universe \cite{Amendola:1993uh,Sushkov:2009hk,Germani:2010hd,Saridakis:2010mf,Huang:2014awa,Yang:2015pga,Koutsoumbas:2013boa}. The applicability of Horndeski theories at cosmological scales can be tested from the measurements of the propagation speed of Gravitational Waves (GWs). If a scalar field is coupled to the Einstein tensor, then the propagation speed of GWs differs from the speed of light \cite{Germani:2010gm,Germani:2011ua} constraining in this way the derivative coupling parameter of the Horndeski theories at cosmological scales \cite{Lombriser:2015sxa,Lombriser:2016yzn,Bettoni:2016mij,Baker:2017hug,Creminelli:2017sry,Sakstein:2017xjx,Ezquiaga:2017ekz}. Therefore, precise measurements of the propagation speed of GWs are very useful tools for constraining the applicability of the Horndeski theory \cite{Gong:2017kim}.

In modified gravity theories by exploring the geodesics of particle motion around a black hole background and solving the geodesic equations we can reveal the properties of these compact objects. Also, observations within the Solar System, such as light deflection, the perihelion shift of planets, and gravitational time delay, can be investigated by the study of the geodesics for particle motion around a black hole background. The motion of charged particles in the Reissner-Nordstr\"om spacetime has been discussed in \cite{Olivares:2011xb}. Geodesics of the magnetically charged Garfinkle-Horowitz-Strominger stringy BH \cite{Garfinkle:1990qj} have been analyzed in \cite{Kuniyal:2015uta, Soroushfar:2016yea},  and the motion of massive particles with electric and magnetic charges was studied in Ref. \cite{Gonzalez:2017kxt}. Studies have calculated the perihelion precession of planetary orbits and the bending angle of null geodesics for various gravity theories in string-inspired models \cite{Chakraborty:2012sd}. The effects of gravity on the Solar System have also been studied in black hole AdS geometries by analyzing the motion of particles in AdS spacetime \cite{Cruz:2004ts,Vasudevan:2005js, Hackmann:2008zz,Hackmann:2008zza, Olivares:2011xb,Cruz:2011yr,Larranaga:2011fp,Villanueva:2013zta,Gonzalez:2013aca,Gonzalez:2015jna}.

In this work, we consider the motion of a test massive particle around the horizon of asymptotically flat Hordenski black holes and we study the generated geodesic structure. The motion of the photons around these black holes has been studied  and three classical tests of gravity in the Solar System, such as the bending of the light, the gravitational redshift, and the Shapiro time delay were considered in order to constraint the coupling parameters of the scalar field to gravity \cite{Carvajal:2025emj}. We will consider a subclass of the Horndeski theory with a scalar field coupled kinetically to the Einstein tensor, and we will work with a modified black hole solution presented in \cite{Babichev:2017guv}. In these black hole solutions of the Horndeski theory the derivative coupling of the scalar field to Einstein tensor appears in the metric function of the black hole as a primary charge. The purpose of this work is to complete the analysis of the geodesic structure and to constrain the parameters appearing in the metric function using Solar System observations, in order to test the viability of the Horndeski theory.

The paper is organized as follows. In Sec. \ref{bhHorn}, we briefly review the four-dimensional Horndeski black hole solution introduced in Ref. \cite{Babichev:2017guv}, which serves as the background geometry for our analysis. In Sec. \ref{geodesic}, we present the equations of motion and analyze the corresponding effective potential. Analytical solutions to the equations of motion are obtained in Sec. \ref{L0} for the case of nonvanishing angular momentum, and in Sec. \ref{LN0} for the case of vanishing angular momentum, thereby establishing the geodesic structure of the spacetime. Finally, in Sec. \ref{conclusion}, we summarize our results and present concluding remarks.

\section{Four-Dimensional Horndeski Black Hole}
\label{bhHorn}

\begin{figure}[H] 
    \begin{center}
        \includegraphics[width=9cm]{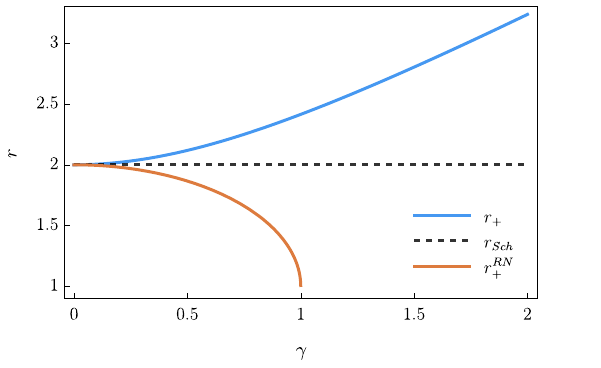}
    \end{center}
    \caption{The first light blue solid curve represents the event horizon of the Horndeski black hole, the second black dashed line corresponds to the Schwarzschild black hole, and the third orange solid curve represents the external event horizon of the Reissner--Nordstr\"{o}m black hole, with $M=1$ in all three cases. Furthermore, for the last curve we set $Q = \gamma$ in $r_+^{RN}$.}
    \label{feventgamma}
\end{figure}

In this section, after reviewing the Horndeski theory, we will discuss a particular hairy black hole solution {\bf{\cite{Babichev:2017guv}}} of the Horndeski theory generated by a scalar field non-minimally coupled to the Einstein tensor. The action of the Horndeski theory \cite{Horndeski:1974wa} is given by
\begin{equation}
\label{acth}
S=\int d^4x\sqrt{-g}(L_2+L_3+L_4+L_5-2\Lambda)~,
\end{equation}
where 
\begin{eqnarray}
    L_2&=&G_2(\phi,X), \\
    L_3 &=&-G_3(\phi,X)\Box \phi,\\
    L_4 &=&G_4(\phi,X)\mathcal{R} + G_{4,X} [(\Box\phi)^2 \nonumber \\
    & & -(\nabla_\mu\nabla_\nu\phi)(\nabla^\mu\nabla^\nu\phi) ]
\end{eqnarray}

and
\begin{eqnarray}
\notag L_5&=&G_5(\phi,X)G_{\mu\nu}\nabla^\mu\nabla^\nu\phi-\frac{1}{6}G_{5,X}[(\Box\phi)^3 \\
 &&  -3(\Box\phi)(\nabla_\mu\nabla_\nu\phi)(\nabla^\mu\nabla^\nu\phi)\\
&& \notag +2(\nabla^\mu\nabla_\alpha\phi)(\nabla^\alpha\nabla_\beta\phi)(\nabla^\beta\nabla_\mu\phi)]~.
\end{eqnarray}
Here $g=\det(g_{\mu \nu})$ with $g_{\mu \nu}$ the metric tensor, $\mathcal{R}$ and $G_{\mu\nu}$ denote the Ricci scalar and the Einstein tensor respectively and $\Lambda$ is the cosmological constant. The functions $G_i$ with $i=2,3,4,5$ are arbitrary functions of the scalar field $\phi$ and the kinetic term $X = - \frac{1}{2} \partial_{\mu} \phi \partial^{\mu} \phi$, while $\Box\phi=\nabla_\mu\nabla^\mu\phi$ is the scalar field in the d'Alembertian operator with the covariant derivative $\nabla_{\mu}$, further $G_{j,X} = \partial G_j/\partial X$ with $j=4,5$.
By considering the particular case 
\begin{eqnarray}
    G_{2} &=& \eta X, \\
    G_{4} &=& \zeta + \beta \sqrt{-X},\\
    G_{3} &=&0=G_{5},
\end{eqnarray}
where $\eta$ and $\beta$ are dimensionless positive parameters and $\zeta=1/2$, the action can be written as 
\begin{eqnarray}
 \notag   S &=& \int d^4 x \sqrt{-g} \Bigg(\left[\frac{1}{2}+\beta\sqrt{(\partial \phi)^2/2} \right]\mathcal{R}-\frac{\eta}{2}(\partial\phi)^2\\
&&    -\frac{\beta}{\sqrt{2(\partial\phi)^2}} \left[(\Box\phi)^2-(\nabla_\mu\nabla_\nu \phi)^2 \right]-2\Lambda \Bigg) \,.
\end{eqnarray}
An asymptotically flat ($\Lambda=0$) Horndeski black hole  
solution \cite{Babichev:2017guv} is given by
\begin{equation}
\label{metric}
 ds^{2}=-f(r)\, dt^{2}+\frac{ dr^{2}}{f(r)}+r^{2}( d\theta^{2}+\sin^{2}\theta\, d\phi^{2})\,,
 \end{equation} where $f(r)$ is the lapse function
\begin{equation} 
    f(r)=1-\frac{2M}{r}-\frac{\gamma ^{2}}{r^{2}}
    \quad \text{with} \quad\gamma = \frac{\beta}{\sqrt{\eta}}  \,\label{1.2}.
\end{equation}
The spacetime allows a unique horizon (the event horizon $r_{+}$), which is located at
	\begin{eqnarray} \label{eventhorizon}
	r_{+}&=&M\left(  1+\sqrt {1+{\gamma^2\over M^2} }\right) ,
	\label{g2.1} \\
	\rho_{2}&=&M\left(  1-\sqrt {1+{\gamma ^2\over M^2} }\right) ,
	\label{g2.1b}
	\end{eqnarray}
	where  $	\rho_{2}$ is a negative solution. The event horizon $r_{+}$ given by \eqref{eventhorizon} increases as $\gamma > 0$ increases, as shown in Fig.~\ref{feventgamma}, this contrasts with the Schwarzschild and Reissner–Nordstr\"om cases, whose event horizons are $r_{s} = 2M$ and $r_{+}^{RN} = M + \sqrt{M^{2} - Q^{2}}$, respectively. In the latter case, the event horizon decreases as $Q^{2}$ increases.

\section{The Motion Equations and The Effective Potential}
\label{geodesic}

In order to study the motion of test particles in the background (\ref{metric}), we use the standard Lagrangian approach \cite{Chandrasekhar:579245,Cruz:2004ts,Villanueva:2018kem}.
The corresponding Lagrangian is
\begin{equation}
2\mathcal{L} =-f(r)\, \dot{t}^{2}+%
\frac{\dot{r}^{2}}{ f(r) }%
+r^{2}\left( \dot{\theta}^{2}+\sin ^{2}\theta \,\dot{\phi}^{2}\right) =-m^2\,.
\label{g6}
\end{equation}%
Here, $m=1$ is the mass of the particle and the dot refers to the derivative with respect to the proper time $\tau$. Since $(t, \phi)$ are cyclic coordinates, their corresponding
conjugate momenta $(\Pi _{t}, \Pi _{\phi })$ are conserved
and given by
\begin{eqnarray}\label{g7}
\Pi _{t}&=&-f(r)\, \dot{t} \equiv -E\,, \\
\Pi _{\phi }&=&r^{2}\sin ^{2}(\theta) \,\dot{\phi} \equiv L\,,
\end{eqnarray}%
where $E$ and $L$ correspond to the energy and angular momentum for the massive particle, respectively. The motion is performed in an invariant plane, which we fix at $\theta =\pi/2$, so we obtain the
following expressions
\begin{equation}\label{g8}
    \dot{t}=\frac{E}{f(r) }\,,%
    \quad \text{and} \quad \dot{\phi}=\frac{L}{r^{2}}\,.
\end{equation}
These relations together with Eq.(\ref{g6}) allow us to obtain the following
differential equations
\begin{eqnarray}\label{g9}
    \left(\frac{dr}{d\tau}\right)^{2}&=&E^2 - V_{\text{eff}}^2(r) \,,\\ \label{g10}
    \left(\frac{dr}{dt}\right)^{2}&=&\frac{f(r)^2}{E^2}
    \,\left(E^2 - V_{\text{eff}}^2(r) \right)\,, \\ \label{g11}
    \left(\frac{dr}{d\phi}\right)^{2}&=&\frac{r^{4}}{L^2}\left( E^2-V_{\text{eff}}^2(r) \right)\,,
\end{eqnarray}%
where the effective potential $V_{\text{eff}}^2 (r) $, is defined as
\begin{eqnarray} \label{eph}
V_{\text{eff}}^2 (r) := f(r)
\left(1 +\frac{L^{2}}{r^{2}}\right)\,.  \label{g12}
\end{eqnarray}%

\begin{figure}[H]
    \begin{center}
        \includegraphics[width=9cm]{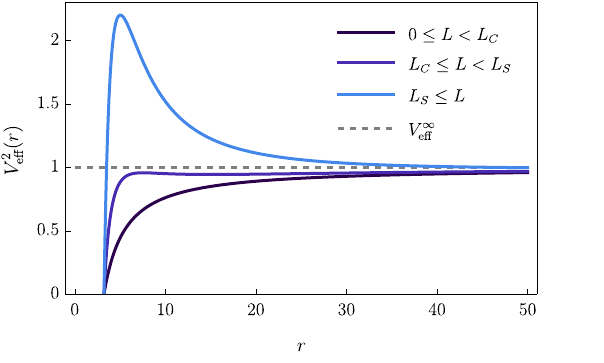}
    \end{center}
    \caption{Three different behaviors of the effective potential are shown for a massive particle. We consider a black hole of mass $M=1$ and $\gamma=2$. The black, purple, and blue curves correspond to $L=0$, $L=5$, and $L=10$, respectively, while the dashed gray line indicates the horizontal asymptotic value $V_{\text{eff}}^{\infty}=1$. Here, $L_C=4.751$ and $L_S=5.330$.}
    \label{fPotential}
\end{figure}

The behavior of the effective potential is shown in Fig.~\ref{fPotential} for massive test particles with three different values of angular momentum. In the figure,
$L_C$ denotes the critical angular momentum beyond which circular orbits are allowed, $L_S$ denotes the angular momentum beyond which the scattering zone exists and $V_{\text{eff}}^\infty := \lim_{r\to \infty}V_{\text{eff}}(r)=1$ the horizontal asymptotic value of the effective potential. For angular momenta in the range $0 \leq L < L_C$, the spacetime allows second kind orbits when $E < 1$: the particle reaches a turning point and then falls into the horizon. When $E \geq 1$, the particle can either escape to infinity or plunge into the horizon. In the range $L_C \leq L < L_S$, the spacetime admits stable and unstable circular orbits, first-kind orbits (such as planetary orbits), critical trajectories, and second kind orbits, for $E < 1$. For $E \geq 1$, the particle may escape to infinity or fall into the black hole. For $L \geq L_S$, the spacetime includes a scattering zone where both first kind and second kind trajectories are possible, as long as $E < V_{\text{eff}}(r_U)$, where $r_U$ denotes the unstable radius for which $V_{\text{eff}}^2$ reaches its maximum value. When $E = V_{\text{eff}}(r_U)$, the allowed motions include unstable circular orbits and critical trajectories. If $E > V_{\text{eff}}(r_U)$, the particle can escape to infinity or plunge into the horizon. In Appendix \ref{CP} and \ref{SP}, we show the analytical solution for the critical and scattering points.

\begin{figure}[H]
    \begin{center}
        \includegraphics[width=9cm]{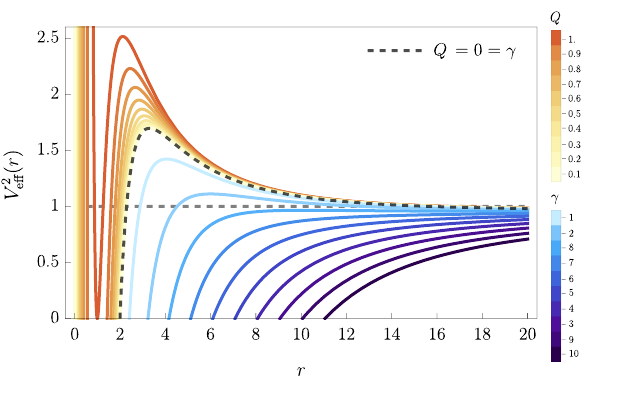}
    \end{center}
    \caption{Effective potentials are shown for a massive particle with angular momentum $L=6$. The black dashed curve corresponds to the Schwarzschild black hole, the curves above it to the Reissner--Nordstr\"{o}m black hole, and the curves below it to the Horndeski black hole, the three cases with $M=1$. The gray horizontal dashed line indicates the asymptotic value of the effective potential.}
    \label{fPotentialAll}
\end{figure}

{\bf{$\gamma$-Analysis.}} The coupling parameter $\gamma$ gives us a lot of physics information by itself through the effective potential (\ref{eph}) and also if it is compared with Schwarzschild and Reissner-Nordstr\"{o}m effective potentials that are given by
\begin{eqnarray}
    V^2_{\text{Sch}}(r) &=& \left(1+\frac{2M}{r} \right) \left(1 + \frac{L^2}{r^2} \right)\,, \\
    V^2_{\text{RN}}(r) &=& \left(1+\frac{2M}{r} + \frac{Q^2}{r^2} \right) \left(1 + \frac{L^2}{r^2} \right)\,,
\end{eqnarray}
where $Q$ is the black hole electric charge (in natural units $c=1=G$). In order to compare, in Fig.~\ref{fPotentialAll}, we plot the potentials for $M=1$, and $L=6$. We observe that $V_{\text{eff}}^2(r) \leq V_{\text{Sch}}^2(r) \leq V_{\text{RN}}^{2}(r)$ for all $r>0$. The Reissner-Nordstr\"{o}m black holes exhibit a higher maximum potential compared to Schwarzschild black holes, resulting in a broader scattering region. In contrast, the Horndeski black holes show a lower maximum potential, leading to a reduced scattering region. This region can even disappear for sufficiently large values of the parameter $\gamma$.

\section{Motion with $L\neq 0$}
\label{L0}

In the following, the trajectories will be classified into two cases; bounded ($E<1$) and unbounded ($E\geq 1$) trajectories.

\subsection{Bounded Orbits}

Using the angular geodesic equation (\ref{g11}), we obtain different types of integral solutions bound to different angular momentum values $L$ and energy values $E$ of the massive particle. If $E<1$, the integral has the form
\begin{eqnarray} \label{angular4}
    \phi(r) = \pm \frac{L}{\sqrt{1-E^2}} \int_{r_0}^{r} \frac{dr'}{\sqrt{P_4(r')}}\,,
\end{eqnarray}
where $r_0$ is the initial radial point and $P_4$ is a quartic polynomial
\begin{eqnarray} \label{quarticpol}
    P_4(r) &:=& -(r^4+a_{0} r^3+b_{0} r^2+c_{0} r+d_{0})\\ 
    & = & -(r-r_1) (r-r_2) (r-r_3) (r-r_{4})\,, \label{Proots}
\end{eqnarray}
with the coefficients
\begin{equation}
    a_0 = \frac{2M}{E^2-1}\,,\quad
    b_0 = \frac{\gamma^2-L^2}{E^2-1}\,,
\end{equation}
    \begin{equation}
    c_0 = \frac{2ML^2}{E^2-1}\,, \quad d_0 = \frac{\gamma^2 L^2}{E^2-1}\,.
\end{equation}
Using the Descartes method to find roots of quartic polynomial, we obtain
\begin{eqnarray}\label{rpaf4} 
r_1&=& +\frac{k}{2} + \frac{1}{2} \sqrt{k^2-4 n_1} - \frac{a_0}{4}  \,, \\
r_2&=& +\frac{k}{2} - \frac{1}{2} \sqrt{k^2-4 n_1} - \frac{a_0}{4}  \,, \\ 
r_3&=&  -\frac{k}{2} + \frac{1}{2} \sqrt{k^2-4 n_2} - \frac{a_0}{4}  \,, \\
r_{4}&=& -\frac{k}{2} - \frac{1}{2} \sqrt{k^2-4 n_2} - \frac{a_0}{4} \,,
\end{eqnarray}
where $k$ satisfies the sixth-order equation
\begin{eqnarray} \label{angulark}
    k^6 + 2 \,p \, k^4+(p^2-4s) \, k^2 - q^2=0,
\end{eqnarray}
with
\begin{equation}
    p = b_0 - \frac{3}{8} a_0^2\,,\quad  
    q = \frac{a_0^3}{8} - \frac{a_0 b_0}{2} + c_0\,,
\end{equation}
    \begin{equation}
    s = d_0- \frac{a_0 c_0}{4}+ \frac{a_0^2 b_0}{16} - \frac{3}{256} a_0^4\,,
 \end{equation}
and 
\begin{eqnarray}
    n_1 = \frac{1}{2} \left( k^2 + p + \frac{q}{k} \right) \,, \quad 
    n_2 = \frac{1}{2} \left( k^2 + p - \frac{q}{k} \right)\,.
\end{eqnarray}
The solution for (\ref{angulark}) is given by 
\begin{eqnarray}
    k &=& \sqrt{2 \sqrt{\frac{\bar{p}}{3}} \sin \left( \frac{\theta}{3} + \frac{2 \pi}{3} \right) - \frac{2 p }{3} }\,,
\end{eqnarray}
where the constants are
\begin{eqnarray}
    \theta &=& \arcsin \left( \frac{\bar{q}}{2} \sqrt{\frac{3^3}{\bar{p}^3}} \right) \,,\\
    \bar{p} &=& \frac{ p ^2}{3} +4s \,,\\
    \bar{q} &=&- \frac{2 p^3}{27} - q^2+ \frac{8 p s }{3}\,.
\end{eqnarray}
If $E=1$, then the angular integral we need to evaluate is
\begin{eqnarray} \label{angular3}
    \phi(r)\; \Big|_{E=1} = \pm \frac{L}{\sqrt{2M}} \int_{r_0}^{r} \frac{dr'}{\sqrt{P_3(r')}}\,,
\end{eqnarray}
where $P_3$ is a cubic polynomial
\begin{eqnarray} \label{threepol}
    P_3(r) &:=& r^3 + a_1 r^2 + b_1 r + c_1 \\
    &=& (r-\bar{r}_1)(r-\bar{r}_2)(r-\bar{r}_3)
\end{eqnarray}
with the coefficients
\begin{equation}
    a_1=\frac{\gamma^2-L^2}{2M}\,, \quad
    b_1= L^2\,, \quad
    c_1= \frac{\gamma^2 L^2}{2M}\,.
\end{equation}
In these cases, the radial roots have the form
\begin{eqnarray}
    \bar{r}_1&=& 2 \sqrt{\frac{p_1}{3}} \sin \left( \frac{\theta_1}{3} \right) - \frac{a_1}{3}\,,\\
    \bar{r}_2&=& 2 \sqrt{\frac{p_1}{3}} \sin \left( \frac{\theta_1}{3} + \frac{2\pi}{3}\right) - \frac{a_1}{3}\,,\\
    \bar{r}_3&=& 2 \sqrt{\frac{p_1}{3}} \sin \left( \frac{\theta_1}{3} + \frac{4\pi}{3}\right) - \frac{a_1}{3}\,,
\end{eqnarray}
where
\begin{equation}
    \theta_1 = \arcsin\left( \frac{q_1}{2} \sqrt{ \left( \frac{3}{p_1} \right)^3} \right)\,, 
\end{equation}

\begin{equation}
    p_1 =  \frac{a_1^2}{3} - b_1\,, \quad
    q_1 = c_1 + \frac{2a_1^3}{3^3} - \frac{a_1 b_1 }{3}\,.
\end{equation}
 If $E>1$, it is more appropriate to extract $\sqrt{E^2-1^2}$ from the denominator of the angular integral, then the polynomial that appears is $Q_{4}(r):=-P_4(r)$. As we mentioned at the beginning of this section, there are many different solutions to the angular integral, which we detail in what follows.

\subsubsection{Circular orbits.}

 The existence of a maximum and a minimum in the potential corresponds to unstable (U)/stable(S) circular orbits, respectively. Thus, throught the radial derivative of the effective potential (\ref{eph}), the equation $V_{\text{eff}}^2\,'(r)=0$ yields

\begin{eqnarray} \label{polyrc}
    r^3+a_cr^2+b_cr+c_c=0\,,
\end{eqnarray}
where
\begin{equation}
    a_c= \frac{\gamma^2 - L^2}{M}\,, \quad 
    b_c=3L^2\,,\quad 
    c_c= \frac{2L^2\gamma^2}{M}\,,
\end{equation}
the index $c$ is for {\it the critical points}. Through the Tschirnhaus transformation $r=x-\frac{a_c}{3}$ and the cubic sine equation, we find the following roots
\begin{eqnarray}
    r_{c,1}&=& 2 \sqrt{\frac{p_c}{3}} \sin \left( \frac{\theta_c}{3} \right) - \frac{a_c}{3}  ,\\
    r_{c,2}&=& 2 \sqrt{\frac{p_c}{3}} \sin \left( \frac{\theta_c}{3} + \frac{2\pi}{3}\right) - \frac{a_c}{3} ,\\
    r_{c,3}&=& 2 \sqrt{\frac{p_c}{3}} \sin \left( \frac{\theta_c}{3} + \frac{4\pi}{3}\right) - \frac{a_c}{3},
\end{eqnarray}
where the angle $\theta_c$ is
\begin{eqnarray}
    \theta_c &=& \arcsin\left( \frac{q_c}{2} \sqrt{ \left( \frac{3}{p_c} \right)^3} \right)\,,
\end{eqnarray}
and the coefficients 
\begin{equation}
    p_c =  \frac{a_c^2}{3} - b_c\,, \quad
    q_c = c_c + \frac{2a_c^3}{3^3} - \frac{a_c b_c }{3}\,,
\end{equation}
where $r_{c,1} =: r_U$ and $r_{c,2} =: r_{S}$ for unstable and stable circular region curves, respectively, and $r_{c,3}$ is a negative root without physical interest.

On the other hand, the proper period in such a circular orbit of radius $r_{c}$ is
    \begin{equation}\label{p1}
	T_{\tau}=2\pi\,r_{c}\, \sqrt{\frac{ r_{c}^2-3 M r_{c}-2 \gamma^2}{M r_{c}+ \gamma^2 }}\,,
	\end{equation}
	and the coordinate period is
	\begin{equation}\label{p2}
	T_t=2\pi\,r_{c}\, \sqrt{\frac{ r_{c}^2}{M r_{c}+ \gamma^2 }}\,.
	\end{equation}	
When the coupling parameter $\gamma$ is null, we recover the periods obtained for Schwarzschild black holes. \\

\begin{figure}[H]
	\begin{center}
		\includegraphics[width=9cm]{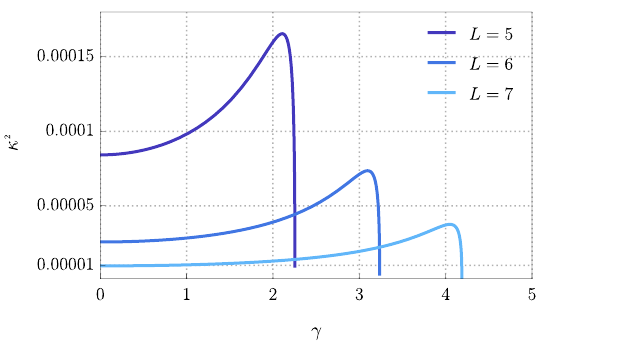}
	\end{center}
	\caption{ The epicycle frequency versus the $\gamma$ parameter with $M=1$ and the angular momenta $L=5$, $L=6$ and $L=7$, respectively. }
	\label{frequency}
\end{figure}

Now, in order to obtain the epicycle frequency for the stable circular orbit, we expand the Taylor series for the effective potential $V_{\text{eff}}(r)$, around the stable circular region $r=r_S$, which yields
\begin{eqnarray}\label{e17}
V_{\text{eff}}^2(r)=V_{\text{eff}}^2 (r_S)+ V_{\text{eff}}^2 \,' (r_S)(r-r_S)\\+{1\over2}V_{\text{eff}}^2\, ''(r_S)(r-r_S)^2+ \cdots\,, \nonumber
\end{eqnarray}
where $(\cdot)\,'$ means derivative with respect to the radial coordinate.
Obviously, in this orbits $V_{\text{eff}}^2 \,'(r_S)=0$. So, by defining the smaller
coordinate $x=r-r_S$, together with the epicycle frequency
$\kappa^2 := V_{\text{eff}}^2 \, ''(r_S)/2 $ \cite{RamosCaro:2011wx},  we can rewrite the above equation as
\begin{equation}\label{e18}
V_{\text{eff}}^2(x)\approx E_S^2+\kappa^2\,x^2\,,
\end{equation}
where ${E_S}:=V_{\text{eff}}(r_S)$ is the energy of the particle at the stable circular orbit.
Moreover, the test particles satisfy the harmonic equation of motion $\ddot{x}=-\kappa^2\,x$, and the epicycle frequency is given by
\begin{equation}\label{e20}
\kappa^{2}= \frac{M\,r_S^{3}-6M^2r_S^{2}-9M\gamma^2r_S-4\gamma^4}{ r_S^4 \left(r_S^2-3 M r_S-2 \gamma^2\right)}\,.
\end{equation}
Notice that $\kappa^2 \rightarrow \kappa_{\text{Sch}}^2$
when $\gamma\rightarrow 0$, where $\kappa_{\text{Sch}}^2$
is the frequency of the epicycle in the Schwarzschild case. In Fig. \ref{frequency} we show the behavior of $\kappa^2$ as a function of $\gamma$ for $L>L_C$. The function $\kappa^2$ exhibits a peak followed by a sharp drop, beyond which the epicycle frequency vanishes, indicating the absence of stable circular orbits for the given values of $L$. Note that for small values of $\gamma$, $\kappa^2>\kappa_{\text{Sch}}^2$, and that $\kappa^2$ increases as $L$ decreases. \\

\subsubsection{Planetary orbits}

\begin{figure}[H]
	\begin{center}
	\includegraphics[width=8cm]{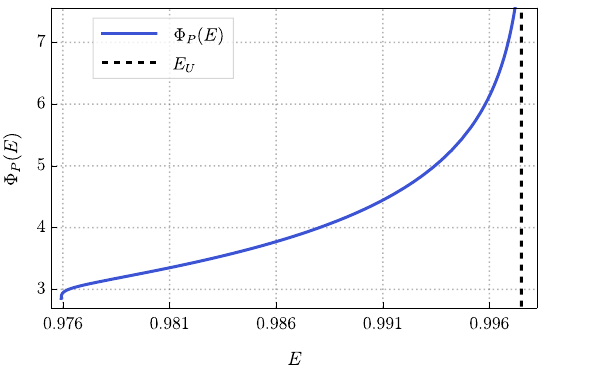}
	\end{center}
	\caption{The behavior of the precession angle $\Phi_P$ for a massive particle, between the energies $E_S<E<E_U$, where  $M=1$, $\gamma=2$ and $L=5.3$.
    }
	\label{fig:anguloalfa1}
\end{figure}

It is possible to find confined orbits of the first kind when $L_C< L < L_S$ with $ E_S < E < E_U < V_{\text{eff}}^{\infty}$ and when $L > L_S$, for energies in the range $ E_S < E < V_{\text{eff}}^{\infty}  < E_U $. This corresponds to energies between the effective potential at the stable circular orbit, $ E_S = V_{\text{eff}}(r_S) $, and that at the unstable circular orbit, $ E_U := V_{\text{eff}}(r_U)$, provided that the latter does not exceed the asymptotic value $ V_{\text{eff}}^{\infty}=1$. In this case, we adopt a specific notation for the radial roots of the polynomial in Eq.~\eqref{quarticpol}: we define $ r_a := r_1$  as the apoastron distance, $r_p := r_2$ as the periastron distance, and $ r_f :=  r_3$ as a turning point corresponding to second-kind trajectories. The root $r_{4}$ is negative and therefore has no physical relevance. For this configuration, we solve the angular integral in Eq.~\eqref{angular4}, obtaining the following solution:

\begin{eqnarray}
    \label{rphiP}
    r_P(\phi) = r_p + \frac{1}{\wp \left(  \kappa_{P} \, \phi + \varphi_{0,P} \right) - A_{P}/3}\,,
\end{eqnarray}
where the index $P$ is for {\it planetary orbits} and $\wp$ corresponds to the $\wp-$Weierstrass elliptic function with the Weierstrass invariants
\begin{equation}
    g_{2,P} = 4 \left( \frac{A_{P}^2}{3} - B_{P} \right)\,, 
\end{equation}
\begin{equation}
    g_{3,P} = 4 \left( \frac{A_{P} B_{P}}{3} - \frac{2A_{P}^3}{27} - C_{P} \right)\,.
\end{equation}
The constants are defined as
\begin{eqnarray}
    A_{P} &=&  w(r_a) + w(r_f) + w(r_{4})  \,,\\
    B_{P} &=&  w(r_{a}) w(r_{f}) + w(r_{a}) w(r_{4})  + w(r_{f})w(r_{4})  \,,\\
    C_{P} &=& w(r_{a}) w(r_{f}) w(r_{4})\,,
\end{eqnarray} 
where the auxiliary function $w(r)$ is defined by $w(r) := 1/(r_2-r)$, with $r_2=r_p$ in this case. The parameters are given by $\kappa_{P} = \sqrt{1-E^2} /(2 L \sqrt{-C_{P} } )$, and
$\varphi_{0,P} = \wp^{-1} \left( A_{P}/3 - w(r_a) \right)$,
where the initial point is located at the apoastron distance $r_0 = r_a$. This solution (\ref{rphiP}) allows us to calculate the precession angle $\Phi_P$, defined as 
\begin{eqnarray}
    \Phi_P:= 2\phi_{(a\to p)}-2\pi,
\end{eqnarray}
where $\phi_{(a\to p)}$ is the angle from the apoastro to the periastro. Thus, it is straightforward to find
\begin{eqnarray}
\label{TRN}
    \Phi_P&=&\frac{2}{\kappa_P}\wp^{-1} \left(A_P/3 - w(r_a)\right)-2\pi\,.
\end{eqnarray}
This exact solution for the angle of precession depends on the spacetime parameters and the particle motion constants. So, according to Eq. (\ref{TRN}), it is possible to see the behavior of $\Phi_P$ as a function of the energy $E$, that is, $\Phi_P \equiv \Phi_P (E) $, as we see in Fig.~\ref{fig:anguloalfa1}. The angle of precession
increases from $E_S$ when the energy increases, and this angle diverges when $E\rightarrow E_U$.

\begin{figure}[H]
	\begin{center}
		\includegraphics[width=9cm]{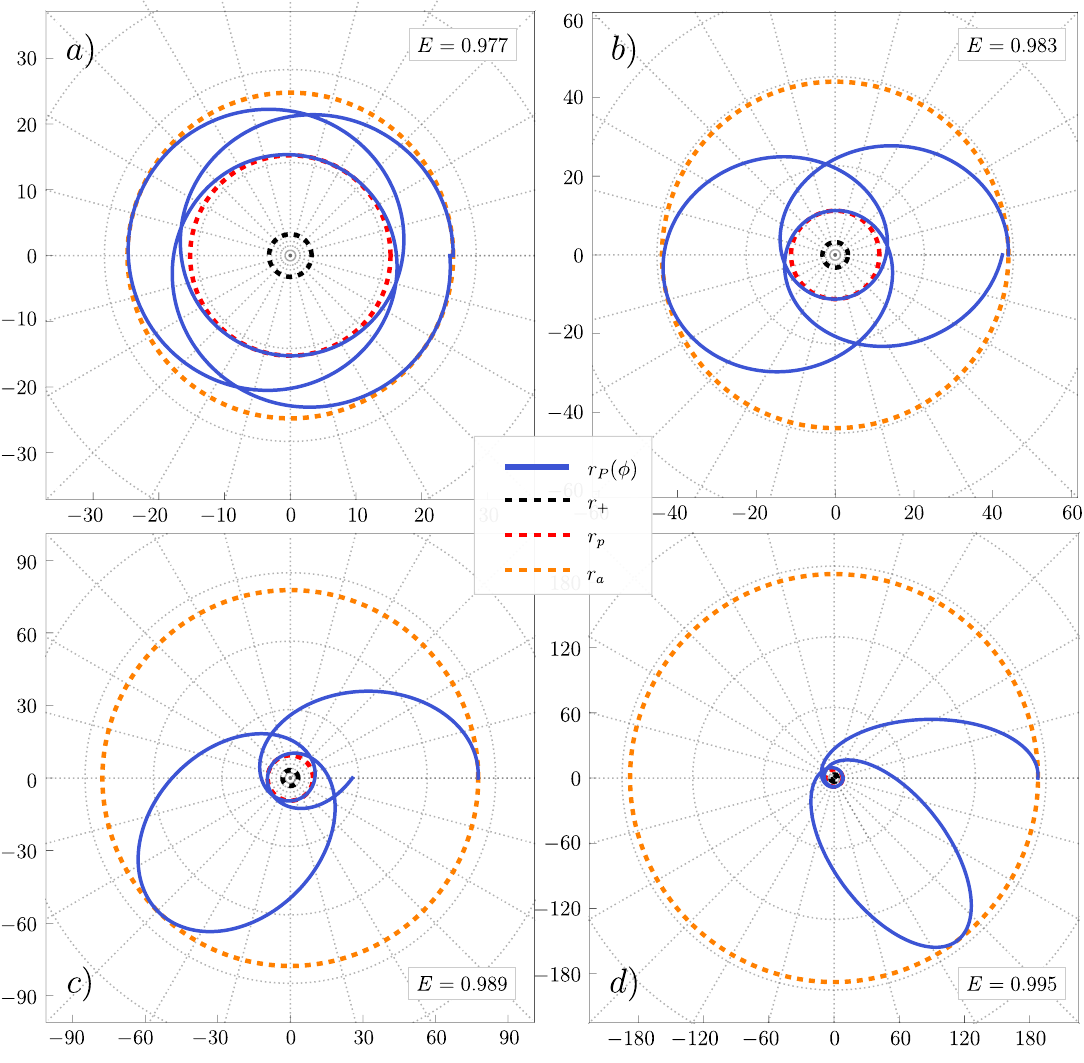}
	\end{center}
	\caption{Planetary orbit $r_P(\phi)$ represented by the solid blue curve for a massive particle; with $\gamma=2$, $M=1$ and $L=5.3$. The orange dashed circle represents the apoastro distance $r_a$, the red dashed circle the periastro $r_p$ and the black dashed circle the event horizon $r_+$; a) is for energy $E=0.977$, b) $E=0.983$, c) $E=0.989$ and d) $E=0.995$.
    }
	\label{fig:orbitasplanetarias}
\end{figure}

For a better visualization of how the precession angle increases when the energy increases, we show such
behavior 
in Fig.~\ref{fig:orbitasplanetarias}, where a particle with fixed angular momentum is taken, but with four different energies.

{\bf{Perihelion precession.}}   It should be noted that, unlike in General Relativity, a general Birkhoff’s theorem does not necessarily hold in Horndeski gravity. Therefore, in this work we explicitly assume that the vacuum solution employed provides an accurate description of the exterior gravitational field of a spherically symmetric source such as the Sun, in analogy with the Schwarzschild solution in GR \cite{Chandrasekhar:579245}. This assumption is expected to remain valid in the weak-field regime, where deviations from GR are small. So, to compute the perihelion shift in a simple and transparent way, we first establish a connection with Kepler's classical problem. Starting from Eq.~\eqref{g11}, and performing the standard change of variables $u=1/r$, we obtain
\begin{equation}
\label{Binet0}
    (u')^2=\frac{E^2-1}{L^2}-u^2+\frac{2M}{L^2}u+2Mu^3+\frac{\gamma^2}{L^2}u^2+\gamma^2u^4\,.
\end{equation}
where $u'=du/d\phi$. Now, we convert the first-order Eq.~\eqref{Binet0}  into a second-order  equation by differentiation with respect to $\phi$, and setting the common factor $u'$, we find
\begin{equation}
\label{Binet}
    u''+u=\frac{M}{L^2}+3Mu^2+\frac{\gamma^2}{L^2}u +2\gamma^2u^3\,.
\end{equation}
This is the Binet equation for Horndeski spacetime and exhibits a structure analogous to that of the classical Kepler problem. Now, the following treatment, performed by Adler, Bazin, and Schiffer \cite{Adler},
allows us to derive the formula for the advance of the
perihelia of planetary orbits.
The starting point is to
consider the Binet equation ~\eqref{Binet}. By defining $A=M/L^2$ and introducing the small dimensionless quantity $\epsilon=3MA$, Eq. (\ref{Binet}) can be rewritten as

\begin{equation}
\label{Binet2}
    u''+u=A+\frac{\epsilon}{A}u^2+\frac{\gamma^2\epsilon}{3M^2}u+\frac{2\gamma^2\epsilon}{3MA}u^3\,.
\end{equation}
This can be solved by assuming the solution ansatz $u(\phi)\approx u_0(\phi)+\epsilon v(\phi)+\mathcal{O}(\epsilon^2)$. So, substituting in the differential equation (\ref{Binet2}), we obtain
\begin{equation}
\label{Binet3}
u_0''+u_0+\epsilon(v''+v)=A+\frac{\epsilon}{A}u_0^2+\frac{\gamma^2\epsilon}{3M^2}u_0+\frac{2\gamma^2\epsilon}{3MA}u_0^3+\mathcal{O}(\epsilon^2)\,.
\end{equation}
Equating the zeroth-order terms in $\epsilon$, yields
\begin{equation}
    u_0''+u_0=A\,.
\end{equation}
The solution for $u_0$ is 
\begin{equation}
\label{B0}
    u_0=A+B\cos(\phi+\delta)\,=\frac{1}{\ell}+\frac{e}{\ell}\cos\phi\,,
\end{equation}
where appropriate axes orientations have been performed given $\delta=0$,  $A=1/\ell$, and $B=e/\ell$, being $\ell$ the latus rectum and $e$ the eccentricity of the Keplerian elipse. 

Now, equating the first-order $\epsilon$ terms in Eq. (\ref{Binet3}), yields

\begin{equation}
\label{v}
    v''+v= \frac{u_0^2}{A}+\frac{\gamma^2}{3M^2}u_0+\frac{2\gamma^2}{3MA}u_0^3\,.
\end{equation}
Thus, substituting Eq. (\ref{B0}) in Eq. (\ref{v}), we obtain  
\begin{equation} 
\label{m0}
v''+v=\bar{A}+\bar{B}\cos\phi+\bar{C}\cos^2\phi+\bar{D}\cos^3\phi\,,
\end{equation}
where
\begin{equation}
    \bar{A}=\frac{1}{\ell}+\frac{\gamma^2}{3M^2\ell}+\frac{2\gamma^2}{3M\ell^2}\,,\,\, \bar{B}= \frac{2e}{\ell}+\frac{\gamma^2e}{3M^2\ell}+\frac{2\gamma^2e}{M\ell^2}\,,
\end{equation}
\begin{equation}
    \bar{C}=\frac{e^2}{\ell}+\frac{2\gamma^2 e^2}{M\ell^2}\,,\quad \bar{D}= \frac{2\gamma^2  e^3}{3M \ell^2}\,.
\end{equation}
Now, using the trigonometric identities $2\cos^2\phi=1+\cos2\phi$, and $4\cos^3\phi=3+\cos 3\phi$ allows us to write Eq. (\ref{m0}) as 
\begin{equation}  
\label{m1}
v''+v=\tilde{A}+\tilde{B}\cos\phi+\tilde{C}\cos 2\phi+\tilde{D}\cos 3\phi\,,
\end{equation}
where
\begin{equation}
    \tilde{A}=\bar{A}+\frac{\bar{C}}{2}\,,\,\, \tilde{B}= \bar{B}+\frac{3\bar{D}}{4}\,,
\end{equation}
\begin{equation}
    \tilde{C}=\frac{\bar{C}}{2}\,,\,\, \tilde{D}=\frac{\bar{D}}{4}\,.
\end{equation}
The solution to Eq. (\ref{m1}) can be written as $v=v_A+v_B+v_C+v_D$ with
\begin{equation}
    v_A''+v_A=\tilde{A}\,,\,\, v_B''+v_B=\tilde{B}\cos\phi\,,
\end{equation}
\begin{equation}
    v_C''+v_C=\tilde{C}\cos2\phi\,,\,\, v_D''+v_D=\tilde{D}\cos3\phi\,,
\end{equation}
whose solutions are 
\begin{equation}
    v_A=\tilde{A}\,,\,\, v_B=\frac{\tilde{B}}{2}\phi\sin\phi\,,
\end{equation}
\begin{equation}
    v_C=-\frac{\tilde{C}}{3}\cos2\phi\,,\,\, v_D=-\frac{\tilde{D}}{8}\cos3\phi\,.
\end{equation}
Therefore, the solution is 
\begin{eqnarray}
\notag        u&=&\frac{1}{\ell}+\frac{e}{\ell}\cos\phi+\epsilon\frac{\tilde{B}}{2}\phi\sin\phi+\\
   && +\epsilon\left(\tilde{A}-\frac{\tilde{C}}{3}\cos2\phi-\frac{\tilde{D}}{8}\cos 3\phi\right)\,.
\end{eqnarray}
Now, using the following trigonometric identity
$\cos(\phi-\tilde{\epsilon}\phi)\approx \cos(\phi)+\tilde{\epsilon}\phi \sin(\phi)$, allows rewriting the solution as
\begin{equation}
\label{uff}
    u=\frac{1}{\ell}+\frac{e}{\ell}\cos(\phi-\tilde{\epsilon}\phi)+\epsilon\left(\tilde{A}-\frac{\tilde{C}}{3}\cos2\phi-\frac{\tilde{D}}{8}\cos 3\phi\right)\,,
\end{equation}
where
\begin{equation}
    \tilde{\epsilon}=\frac{\epsilon\tilde{B}\ell }{2e}=\frac{3M}{\ell}+\frac{\gamma^2}{2M\ell}+\frac{3\gamma^2}{\ell^2}+\frac{3\gamma^2 e^2}{4\ell^2 }\,.
\end{equation}

In this form, the influence of the various terms on the orbit becomes clear. The fundamental elliptical orbit is described by Eq. (\ref{B0}). The last term introduces small periodic variations in the planet’s radial distance. These variations are subtle and difficult to detect, and because they are periodic, they do not affect the precession of the perihelion. However, the parameter $\tilde{\epsilon}\phi$
  in the cosine argument introduces a non-periodic component. Since $\phi$	
  can become significant, its effect is not negligible. Thus, Eq. (\ref{uff}) can be written in the form
\begin{equation}
\label{uf}
    u=\frac{1}{\ell}+\frac{e}{\ell}\cos(\phi-\tilde{\epsilon}\phi)+\left(\text{periodic terms of order $\epsilon$}\right)\,.
\end{equation}
The perihelion of a planet occurs when $r$ is a minimum or when $u=1/r$  is a maximum.  From Eq. (\ref{uf}) we see that $u$ is maximum when
\begin{eqnarray}
\phi(1-\tilde{\epsilon})=2\pi n\,.
\end{eqnarray}
or approximately
\begin{eqnarray}
\phi = 2\pi n(1+\tilde{\epsilon})=:\phi_n\,.
\end{eqnarray}
Therefore, successive perihelia will occur at intervals of $\Delta\tilde{\phi}=\phi_{n+1}-\phi_{n}= 2\pi (1+\tilde{\epsilon})= 2\pi+\delta\phi$,
where $\delta\phi=2\pi\tilde{\epsilon}$ and it is equal to
\begin{equation}
    \delta \phi=\frac{6\pi M}{\ell}+\frac{6\pi \gamma^2}{\ell^2}\,,
\end{equation}
which correspond to the corrections due to general relativity and Horndeski's theory, respectively, measured in radians per revolution. Here, we have neglected the terms of the eccentricity of second order $e^2$ and the cross term $\gamma^2/M$.
To test the above relation in the solar system, we consider $M = M_{\astrosun} = 1476.1 \,m$, and therefore, the advance of perihelion in arcseconds per century (arcsec/Julian-century), is obtained as
\begin{eqnarray}
    \delta\tilde{\phi} &=& 100{3600\cdot180\over \pi}f_{rev}\left( \frac{6\pi M_{\astrosun}}{\ell}+\frac{6\pi \gamma^2}{\ell^2} \right)\,,\\
    &=&3888\times10^{5}f_{rev}\left( \frac{ M_{\astrosun}}{\ell}+\frac{ \gamma^2}{\ell^2} \right)\,,
     \label{pp}
\end{eqnarray}
in which $f_{rev}$, the revolution frequency, corresponds to the number of orbits per year.
Therefore, the perihelion advance has the standard value
of GR plus the correction term coming from the Horndeski's theory. Considering Eq. (\ref{pp}) we constraint the coupling parameters of the scalar field to gravity $\gamma$, for Mercury, Venus and the Earth, see Table \ref{gamma}. For the case of Venus, the gamma calculation was with the maximum value of the precession $13.2$ (arcsec/Julian-century).
Then, the parameter $\gamma$ of Horndeski's theory, giving the constraint $\gamma\leq 2\times 10^{5}\,m$.
This leads to a more accurate correspondence between the observational data and the theoretical prediction of the perihelion precession within the Solar System.

    \begin{table}[H]
    \begin{center}
	\begin{tabular}{ |p{1.3cm}|p{1.7cm}|p{1.5cm}|p{1.6cm}|p{1.4cm}|  }
		\hline
		Planet&Observed 
        shift.
        $\quad$Seconds per century& Latus rectum ($\ell$)
        ($\times 10^{10}\,m$) & Revolution frequency ($f_{rev}$).
        Orbits 
        per annum& $\qquad \gamma$
        ($\times 10^{6}\,m$)\\
		\hline
		Mercury & $43.11\pm 0.45$   &$5.53$& $4.15$& $0.2$ \\
		\hline
		Venus& $8.4\pm 4.8$   &$10.8$& $1.622$  & $9.2$\\
		\hline
        Earth& $5\pm 1.2$   &$14.9$& $1$  & $8.1$\\
        	\hline
	\end{tabular}
     \caption{Constraint on the  parameter $\gamma$ of the Horndeski's theory.}
  \label{gamma}
  \end{center}
\end{table}

\subsubsection{Second kind trajectories} 

The particles follow second-kind trajectories when they originate at some point in spacetime and subsequently plunge into the horizon; these orbits appear when the energy lies at $0<E<1$. It is important to note that within this energy range and for different values of the angular momentum, multiple solutions may exist. These solutions can be expressed in terms of elementary functions or elliptic functions, depending on the specific configuration.\\

\begin{figure}[H]
	\begin{center}
		\includegraphics[width=8.5cm]{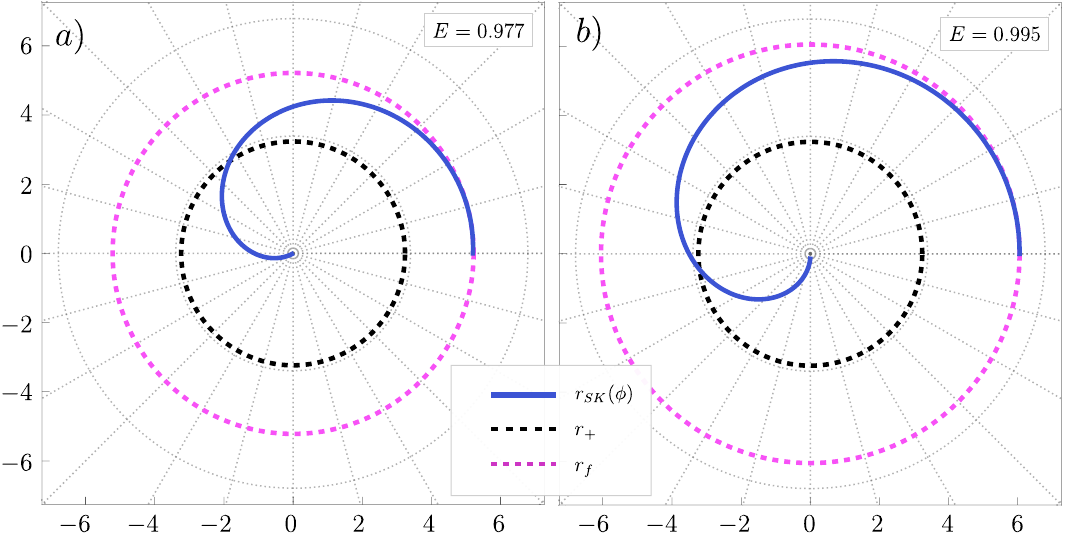}
	\end{center}
	\caption{Behavior of second kind trajectories for a massive particle, represented by the solid blue curve, with $\gamma=2$, $M=1$ and $L=5.3$. The pink dashed circle represents the turning point $r_f$, while the black dashed circle the event horizon $r_+$; a) is for energy $E=E_S=0.977$ and b) $E=0.995$.  }
	\label{fig:secondkind}
\end{figure}

Second kind trajectories for $L>L_C$, and $E=E_S$. An analytical solution for second kind trajectories with stable energy (SKS), namely on $E=E_S$, is described by elementary functions, given by
\begin{eqnarray}
    r_{\text{SKS}}(\phi)&=& r_S + \frac{2(r_S-r_f)(r_S-r_4)}{(r_f-r_4)\sin( \kappa_{\text{SKS}}\, \phi-{\pi\over2})-\rho_{\text{SKS}}}\,,\nonumber \\
    &&
\end{eqnarray}
where
\begin{eqnarray}
    \kappa_{\text{SKS}} &=& \frac{1}{L} \sqrt{(1-E_S^2)(r_S-r_f)(r_S-r_4)}\,,\\
    \rho_{\text{SKS}}&=&2r_S-r_f-r_4\,,
\end{eqnarray}
and the initial point at $r_0 = r_f$.  Fig.~\ref{fig:secondkind}(a), shows the behavior of this type of orbit. \\

Second kind trajectories for $L>L_C$, and $E_S<E <E_U$. For these cases, we use the same solution from $ \eqref{rphiP} $, but we replace the initial condition $ r_0 = r_a $ in the function $ w(r)=1/(r_2-r) $ with $ r_2 = r_f $, and denote the new solution as $ r_{\mathrm{SK}}(\phi) $, where (SK) refers to \textit{second-kind orbits}, see Figure~\ref{fig:secondkind}(b).

\subsubsection{Critical trajectories}

\begin{figure}[H]
	\begin{center}
		\includegraphics[width=6.45cm]{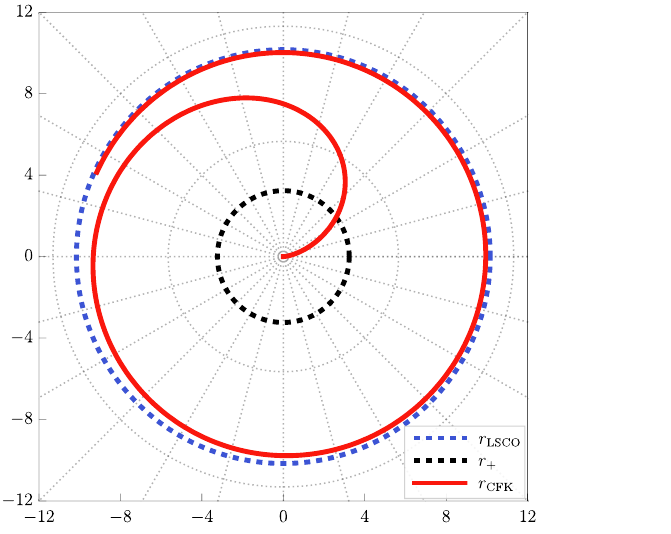}
	\end{center}
	\caption{ The behavior of critical trajectories with  $M=1$, $\gamma=2$, and $L_C=4.75$. Red line for CSK with $E^2=E_{\text{ISCO}}^2=0.932$.
	The small circle represents to  the horizon, $r_+$, and the bigger circle represents to $r_{\text{ISCO}}=10.162$.   }
	\label{fig:criticalLSCO}
\end{figure}

There are two types of trajectories depending on the values of $L$. The first corresponds to the innermost stable circular orbit (ISCO) when $L=L_C$, and the second corresponds to $L_C<L<L_S$. Bounded critical trajectories with $L=L_C$ and $E=E_{\text{ISCO}}$, are given by the following equation of motion 
\begin{equation}
r_{\text{CSK}}(\phi)={r_{\text{ISCO}}^3\over r_{\text{ISCO}}^2+4L_C / \phi^2}\,,
\label{criticaLSCO}
\end{equation}
which we plot in Fig. \ref{fig:criticalLSCO}. This is a second kind trajectory that approach asymptotically to the ISCO.

\begin{figure}[H]
	\begin{center}
		\includegraphics[width=6cm]{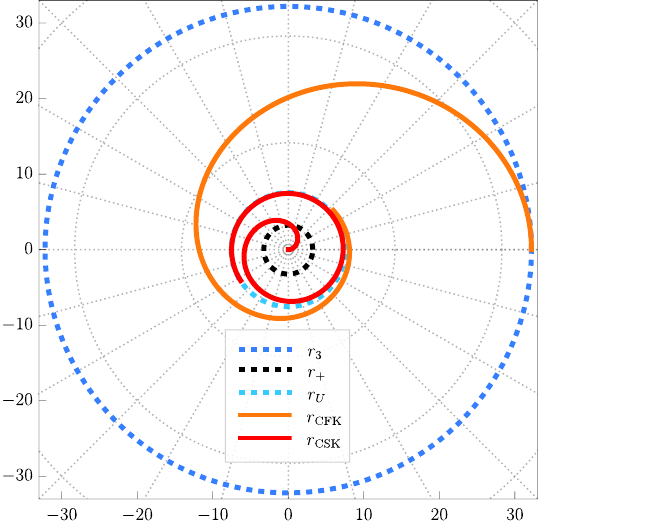}
	\end{center}
	\caption{ The behavior of critical trajectories with  $M=1$, $\gamma=2$, and $L=5$. Orange curve for CFK and red curve for CSK trajectories with $E^2=E_U^2=0.957$, and $r_U=7.544$.
	The small black circle represents to the horizon $r_+$, the medium cyan circle represents to $r_U$, and the bigger blue circle represents to $r_3$.   }
	\label{fig:criticalbounded}
\end{figure}

On the other hand, bounded critical trajectories with $L_C<L<L_S$ and $E=E_{U}$ are given by the following equations of motion 

\begin{equation}
    r_{\text{CFK}}(\phi)=r_U+\frac{2 \left(r_3-r_U\right) \left(r_U-r_4\right)}{\left(r_3-r_4\right) \cosh (\kappa_U \phi )-h}\,,
    \label{criticas02a} 
\end{equation}
\begin{equation}
    r_{\text{CSK}}(\phi)=r_U-\frac{2 \left(r_3-r_U\right) \left(r_U-r_4\right)}{\left(r_3-r_4\right) \cosh (\kappa_U \phi+\delta_0 )+h}\,,
    \label{criticas02b}
\end{equation}
where  
\begin{eqnarray}
 h &=& r_3+r_4-2 r_U\,,\\
 \notag   \kappa_U &=& \frac{ \sqrt{\left(1-E_U^ 2\right)\left(r_3-r_U\right) \left(r_U-r_4\right)}}{ L}\,, \\
 \notag   \cosh \delta_0&=&{2\left(r_3-r_U\right)\left(r_U-r_4\right)+\left(2r_U-r_3-r_4\right)r_U\over r_U\left(r_3-r_U\right)}\,,
\end{eqnarray}
with $r_3 \equiv \left. r_{3} \right|_{E_U}$ and $r_4 \equiv \left. r_{4} \right|_{E_U}$. In Fig.~\ref{fig:criticalbounded}, we show the behavior of the CFK (orange line) and CSK (red line) trajectories, given by Eq. (\ref{criticas02a}) and Eq. (\ref{criticas02b}), respectively. 

\subsection{Unbounded Orbits}

\subsubsection{Scattering trajectories}

\begin{figure}[H] 
    \begin{center}
        \includegraphics[width=8.2cm]{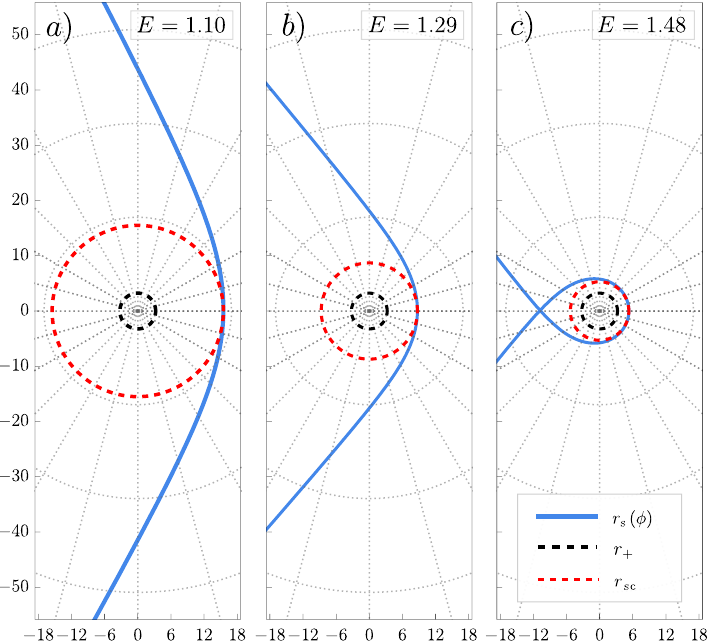}
    \end{center}
    \caption{The scattering trajectories for a massive particle 
    (solid blue curve) 
    are depicted for $M=1$, $\gamma=2$, and $L=10$. The red dashed circle marks the scattering root $r_{sc}$, while the black dashed circle denotes the event horizon $r_{+}$. Panels a), b) and c) show cases with energies $E=1.10$, $E=1.29$ and $E=1.48$, respectively.
}
    \label{fig:Scattering}
\end{figure}

When $ L > L_S $, scattering orbits become possible, provided the energy lies within the range $ 1 < E < E_U $; that is, between the asymptotic value $ V_{\text{eff}}^{\infty} = 1 $ and the energy at the unstable circular orbit $ E_U = V_{\text{eff}}(r_U) $. In this section, we adopt a new notation for the radial roots of the polynomial in Eq.~\eqref{quarticpol}: we define $ r_{\text{sc}} := r_1 $ as the scattering distance, and we redefine $r_{f} := r_2 $ as a turning point associated with second kind trajectories. While the remaining roots, $ r_3 $ and $ r_4 $, are negative and therefore they are not physically relevant. We will use the same solution as in Eq.~\eqref{rphiP}, but with the initial point located at $ r_0 = r_{\text{sc}} $. Consequently, we rename the solution as follows
\begin{eqnarray}
    \label{rphiScatt}
    r_{\text{s}}(\phi) = r_f + \frac{1}{\wp \left(  \kappa_{s} \,\phi + \varphi_{0,s} \right) - A_{s}/3}\,,
\end{eqnarray}
where the index $s$ is for the {\it scattering orbits} and $\wp$ corresponds to the $\wp-$Weierstrass elliptic function with the Weierstrass invariants
\begin{equation}
    g_{2,s} = 4 \left( \frac{A_{s}^2}{3} - B_{s} \right)\,, \quad
    g_{3,s} = 4 \left( \frac{A_{s} B_{s}}{3} - \frac{2A_{s}^3}{27} - C_{s} \right)\,.
\end{equation}
The constants are defined as
\begin{eqnarray}
    A_{s} &=&  w(r_{sc}) + w(r_3) + w(r_{4})  \,,\\
    B_{s} &=&  w(r_{sc}) w(r_{3}) + w(r_{sc}) w(r_{4})  + w(r_{3})w(r_{4})  \,, \\
    C_{s} &=& w(r_{sc}) w(r_{3})w(r_{4})\,,
\end{eqnarray} 
where the auxiliary function $w(r) = 1/(r_2-r)$, with $r_2=r_f$ in this case. The parameters are given by $\kappa_{s} = \sqrt{E^2-1} /(2 L \sqrt{C_s } )$ and
$\varphi_{0,s} = \wp^{-1} \left( A_s/3 - w(r_{sc}) \right)$. In Fig.~\ref{fig:Scattering}, we show the scattering of particles for different energies, where we can observe that as long as the massive particle has a higher energy in the scattering energy range, it is appreciated that the particle manages to surround a larger arc around the black hole.

\begin{figure}[H]
	\begin{center}
		\includegraphics[width=8cm]{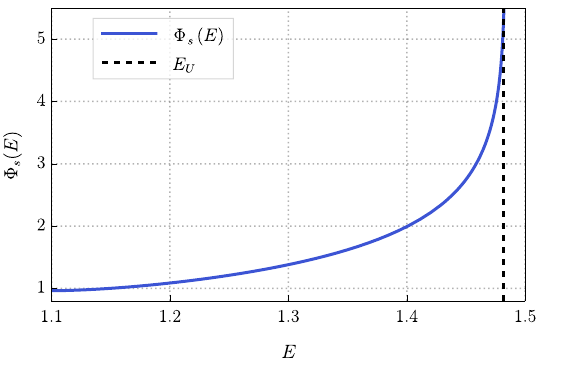}
	\end{center}
	\caption{Scattering angle $\Phi_{s}(E)$ for a massive particle, between the energies $1<E<E_U$, where  $M=1$, $\gamma=2$ and $L=10$.
    }
	\label{fig:scattering}
\end{figure}

The \textit{scattering angle} for particles coming from infinity 
can be written as \cite{Villanueva:2015kua,Fathi:2020sey}
\begin{equation}\label{eq:scatterin_angle0}
    \Phi_s := 2\phi_\infty - \pi,
\end{equation}
 in which $\phi_\infty := \lim_{r\to \infty}\phi(r)$, where $\phi(r)$ is obtained from (\ref{rphiScatt}). Therefore, we have
 \begin{equation}\label{eq:scatterin_angle1}
      \Phi_s ={2\over  \kappa_{s}} \left(\wp^{-1} \left( A_s/3  \right)-\varphi_{0,s} \right)-\pi
 \end{equation}
The value of $\Phi_s$ is directly specified by the initial $E$ and the corresponding particular radial solutions $\bar{r}$, which are determined by the equation $E=V_{\text{eff}}(\bar{r})$. Therefore, these values cannot be considered to evolve in terms of a single variable. However, one can calculate the scattering angle for each particular trajectory, applying Eq.~\eqref{eq:scatterin_angle1}.  Eq.~\eqref{eq:scatterin_angle0} gives the change in the particles' orientation as they approach 
to the black hole at the scattering point $r_{sc}$. To illustrate the corresponding trajectories, we evaluate a set of radial values $r_t$ and compute their associated angular coordinates $\phi(r_t)$, forming the pairs $(r_t, \phi(r_t))$. These are then used to numerically interpolate the trajectory function $r_s(\phi)$. The scattering angles are illustrated in Fig.~\ref{fig:scattering} for particles with different values for $E$. The scattering angle
increases when the energy increases and diverges as the energy approaches the critical value $E\rightarrow E_U$.

Furthermore, the differential angular range of the scattered particles at the angle $\Phi_s$, is given by the solid angle element $\mathrm{d}\Omega = \sin\Phi_s\, \mathrm{d} \Phi_s \mathrm{d}\phi$. In this regard, and by defining the impact parameter $b={L\over E}$, the cross-sectional area of the scattered particles has a differential form $\mathrm{d}\Sigma = b\, \mathrm{d}\phi\, \mathrm{d} b$ \cite{Fathi:2020sey,Fathi:2020sfw}. Therefore, the differential cross section of the scattering is given by
\begin{equation}\label{eq:diffCross-def}
    \Sigma(\Phi_s) := \frac{\mathrm{d}\Sigma}{\mathrm{d}\Omega} = \frac{b}{\sin(\Phi_s)} \left|\frac{\partial b}{\partial\Phi_s}\right|.
\end{equation}
On the other hand, from Eqs.~\eqref{rphiScatt} and \eqref{eq:scatterin_angle0} we have
\begin{equation}\label{eq:Theta_recast}
    \frac{\kappa_s}{2}(\Phi_s+\pi)=\varphi_{1} + \varphi_{2}\,,
\end{equation}
where
\begin{equation}
    \varphi_{1} = \wp^{-1} \left( A_s/3  \right)\,, \quad
     \varphi_{2} =  -\wp^{-1} \left( A_s/3 - w(r_{sc}) \right)\,.
\end{equation}
Thus, by defining
\begin{equation}\label{eq:Psi}
    \Psi(L) := \wp\left(\frac{\kappa_s}{2}(\Phi_s+\pi)
    \right) = \wp\left(
    \varphi_{1} + \varphi_{2}
    \right),
\end{equation}
or equivalently \cite{handbookElliptic}
\begin{equation}\label{eq:Psi(L)}
    \Psi(L) = \frac{1}{4}\left[
    \frac{\wp'(\varphi_{1})-\wp'(\varphi_{2})}{\wp(\varphi_{1})-\wp(\varphi_{2})}
    \right]^2-\wp(\varphi_{1})-\wp(\varphi_{2})\,,
\end{equation}
allows us to rewrite Eq.~\eqref{eq:diffCross-def} as

\begin{eqnarray}
    \Sigma(\Phi_s)&=& b\csc(\Phi_s)
    \left|
    \frac{\partial b}{\partial \Psi}
    \right|\left|
    \frac{\partial \Psi}{\partial \Phi_s}
    \right|  \nonumber\\
    &=& \frac{\kappa_s}{4}\csc(\Phi_s)\left|
    \wp'\left(
    \frac{\kappa_s}{2}(\Phi_s+\pi)
    \right)
    \right|
    \left|
    \frac{\partial  b^2}{\partial \Psi}
    \right|.
   \label{eq:diffCross-def_1}
\end{eqnarray}
Now, using the identity $\frac{\partial  b^2}{\partial \Psi} = \frac{\partial b^2/\partial L}{\partial \Psi/ \partial L}$, the differential cross section of the scattering can be written as
\begin{equation}\label{eq:sigmaTheta_2}
    \Sigma(\Phi_s) = \frac{ \kappa_s\,L}{2E^2}\csc(\Phi_s)\left|
    \wp'\left(
   \frac{\kappa_s}{2}(\Phi_s+\pi)
    \right)
    \right|
    \left|
    \frac{\partial\Psi}{\partial L}
    \right|^{-1}\,.
\end{equation}
The expression for $\Psi$ is analytically cumbersome. However, as in previous cases, the value of Eq.~\eqref{eq:sigmaTheta_2} can be computed numerically for the initial conditions specified that correspond to different scattered trajectories.\\

\subsubsection{Critical trajectories}

\begin{figure}[H]
	\begin{center}
		\includegraphics[width=6cm]{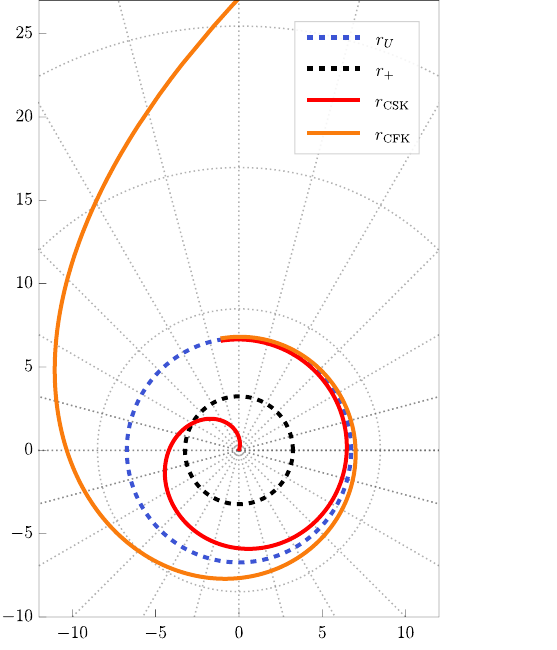}
	\end{center}
	\caption{ The behaviour of critical trajectories with  $M=1$, $\gamma=2$, and $L=5$. Orange line for CFK and red line for CSK trajectories with $E=E_U=1$, and $r_U=7.544$.
	The small black circle represents to the horizon $r_+$,  the medium blue circle represents to $r_1=r_U=6.73$, and the bigger circle represents to $r_3$.   }
	\label{fig:criticalunbounded}
\end{figure}

Unbounded critical trajectories with $L=L_S$ and $E=E_{U}=1$ are given by the following equations of motion 
\begin{eqnarray}
    r_{\text{CFK}}(\phi)&=&r_3+\left(r_1-r_3\right) 
    \left({e^{k_S\phi}+1\over e^{k_S\phi}-1 }\right)^2,
    \label{criticas03a} \\
    r_{\text{CSK}}(\phi)&=&r_3+\left(r_1-r_3\right) 
    \left({e^{k_S\phi}-1\over e^{k_S\phi}+1 }\right)^2,
    \label{criticas03b}
\end{eqnarray}
where  
\begin{eqnarray}
    \kappa_S = \frac{ \sqrt{2M\left(r_1-r_3\right) }}{ L},
\end{eqnarray}
$r_1 \equiv \left. r_{U} \right|_{E_U=1}$, and $r_3\equiv\left. r_{3} \right|_{E_U=1}$. In Fig.~\ref{fig:criticalunbounded}, we show the behavior of the CFK (orange line) and CSK (red line) trajectories, given by Eq. (\ref{criticas03a}) and Eq. (\ref{criticas03b}). 

\begin{figure}[H]
	\begin{center}
		\includegraphics[width=7cm]{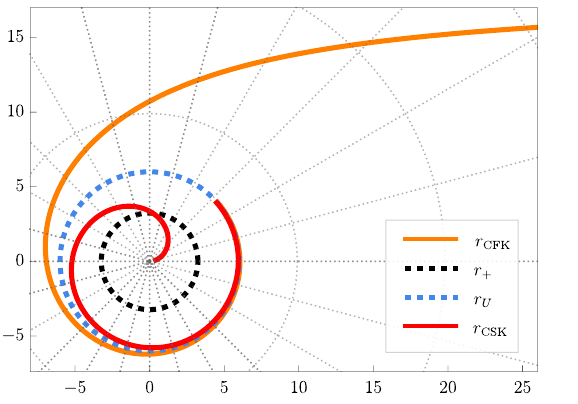}
	\end{center}
	\caption{ The behaviour of critical trajectories with  $M=1$, $\gamma=2$, and $L=5.33$. Orange line for CFK and red line for CSK trajectories with $E=E_U=1.111$, and $r_U=6.0$.
	The small circle represents to the horizon $r_+$, and the bigger circle represents to $r_U$.   }
	\label{fig:criticalUnbounded}
\end{figure}

On the other hand, unbounded critical trajectories with $L>L_{S}$, and $E_{U}>1$ are given by the following equations of motion 
\begin{equation}
r_{\text{CFK}}(\phi)=r_U+\frac{2 \left(r_U-r_3\right) \left(r_U-r_4\right)}{\left(r_3-r_4\right) \cosh (\kappa_u \phi +\varphi_{\infty})+h}\,,
\label{criticas04a} 
\end{equation}
\begin{equation}
r_{\text{CSK}}(\phi) =r_U-\frac{2 \left(r_U-r_3\right) \left(r_U-r_4\right)}{\left(r_3-r_4\right) \cosh (\kappa_u \phi +\varphi_{0})-h}\,,
\label{criticas04b}
\end{equation}
where
\begin{eqnarray}
 h &=& r_3+r_4-2 r_U\,,\\
\notag    \kappa_u &=& \frac{ \sqrt{\left(E_U^ 2-1\right)\left(r_U-r_3\right) \left(r_U-r_4\right)}}{ L}\,, \\
  \notag  \cosh \varphi_{\infty}&=&{2r_U-r_3-r_4\over r_3-r_U}\,, \\
  \notag  \cosh \varphi_{0}&=&{r_3(r_4-r_U)+r_4(r_3-r_U)\over r_U(r_3-r_U)}\,,
\end{eqnarray}
$r_3 \equiv \left. r_{3} \right|_{E_U}$ and $r_4 \equiv \left. r_{4} \right|_{E_U}$. In Fig.~\ref{fig:criticalUnbounded}, we show the behavior of the CFK (orange line) and CSK (red line) trajectories, given by Eq. (\ref{criticas04a}) and Eq. (\ref{criticas04b}) respectively. 

\section{Motion with $L=0$} 
\label{LN0}

The motion of particles with $ L = 0 $ is governed by the condition of vanishing angular momentum, in which case the effective potential reduces to $V_{\text{eff}}^2(r) = f(r)$. In this scenario, the potential lacks a maximum, implying that particles with $ E < 1 $ inevitably fall into the event horizon, whereas those with $ E \geq 1 $ can either escape to spatial infinity or fall into the horizon. The equations describing this type of motion are given by Eqs.~\eqref{g9} and \eqref{g10}, which yield

\begin{eqnarray}
\tau(r) &=&\pm \int^{r}_{r_0}{r'\,dr'\over \sqrt{p_2(r')}}\,, \label{inttau} \\
t(r) &=& \pm \int^{r}_{r_0}{r'^3\,dr'\over(r'-r_+)(r'-\rho_2) \sqrt{p_2(r')}}\,, \label{intt}
\end{eqnarray}
where $p_2$ is the quadratic polynomial defined as
\begin{eqnarray} \label{p2}
\notag p_2(r)&:=&(E^2-1)r^2+2Mr+\gamma^2\,,
\end{eqnarray}
whose roots will give us the return points.

\subsection{Bounded trajectories: $0<E<1$} 
Bounded trajectories are characterized by $E < 1$. 
\begin{eqnarray}
\notag p_{2}(r)&=&(1-E^2)(\tilde{r}_0-r)(r-\tilde{r}_2) \\
&=:& (1-E^2) P_{0}(r)\,.
\end{eqnarray}
Thus, assuming that the particles are placed at $r=\tilde{r}_0$ when $t=0 = \tau$. The turning point $\tilde{r}_0$ is located at 
\begin{eqnarray}
\tilde{r}_{0}(E)&=&{M+\sqrt{M^2+(1-E^2)\gamma^2}\over 1-E^2}\,, 
\end{eqnarray}
and the root $\tilde{r}_2$ is a negative root given by
\begin{eqnarray}
\tilde{r}_{2}(E)&=&{M-\sqrt{M^2+(1-E^2)\gamma^2}\over 1-E^2}\,.
\end{eqnarray}

Therefore, we solve Eq. (\ref{inttau}) and obtain the proper time $\tau(r)$, for the motion of massive particles with $L=0$, in terms of the radial coordinate $r$
\begin{equation}\label{tau1}
\tau \left( r\right) =\sqrt{P_{0}(r)\over1-E^2}+{M\over (1-E^2)^{3/2}}F_0(r)\,,
\end{equation}%
where
\begin{equation}\label{tau2}
F_0(r)=\arcsin\left( {\tilde{r}_0+\tilde{r}_2-2r\over \tilde{r}_0-\tilde{r}_2}\right) + {\pi\over 2}\,.
\end{equation}
Solving Eq. (\ref{intt}), the coordinate time $t$ as a function of $r$ yields
\begin{equation}\label{te1}
t\left( r\right) =E \, \tau \left( r\right)
+{2M\,F_0(r)+A_1\,F_1(r)+A_2\,F_2(r)\over \sqrt{1-E^2}}\,,
\end{equation}
where the functions $F_j(r)$, with $j=1,2$, are given explicitly by
\begin{equation}
  	\label{F0}
F_1(r)={1\over \sqrt{P_{0}(r_+)}}\ln\left|\frac{\tilde{r}_0+\tilde{r}_2-2r_+}
{\tilde{r}_0-\tilde{r}_2}+U_1(r)\right|
\,,\\
\end{equation}
with
\begin{equation}
U_1(r)=\frac{2P_{0}(r_+)+2\sqrt{P_{0}(r_+)P_{0}(r)}}
{(\tilde{r}_0-\tilde{r}_2)(r-r_+)}\,,
  \end{equation}
and
\begin{equation}
  	\label{F0}
F_2(r)={1\over \sqrt{P_{0}(\rho_2)}}\ln\left|\frac{\tilde{r}_0+\tilde{r}_2-2\rho_2}
{\tilde{r}_0-\tilde{r}_2}+U_2(r)\right|
\,,\\
\end{equation}
with
\begin{equation}
U_2(r)=\frac{2P_{0}(\rho_2)+2\sqrt{P_{0}(\rho_2)P_{0}(r)}}
{(\tilde{r}_0-\tilde{r}_2)(r-\rho_2)}\,,
  \end{equation}
where the corresponding constants are
\begin{equation}
A_1={r_+^3\over r_+-\rho_2}\,, \quad \text{and} \quad
A_2=-{\rho_2^3\over r_+-\rho_2}\,.
\end{equation}

\begin{figure}[H]
	\begin{center}
\includegraphics[width=9cm]{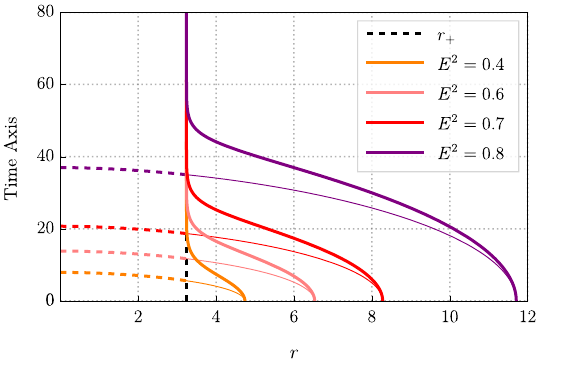}
	\end{center}
	\caption{Proper (thin curve) and coordinate (thick curve) time for the bounded trajectories of test particles with $L=0$. We observe that the return point is greater if the energy increases, as well as, the proper time that take the particles to
cross the event horizon.
	Here, we have used the values   $M=1$, $\gamma=2$, and $E<1$.}
	\label{f2RN}
\end{figure}

Now, in order to visualize the behavior of the proper and coordinate time, we plot in Fig. \ref{f2RN}, their behavior via the solution (\ref{tau1}) and (\ref{te1}), respectively.
We can observe that the particles
cross the event horizon in a finite proper time, but an external observer will see that the particles take an infinite (coordinate) time to do it, the same behavior was observed for the uncharged background.

\subsection{Unbounded trajectories: $E \geq 1$}
The proper time $\tau(r)$ given by Eq. (\ref{inttau}) for the motion of massive particles with $E=1$ and $L=0$ in terms of the radial coordinate $r$ yields

\begin{equation}\label{tau01}
\tau \left( r\right) =\pm{2\left( \tilde{F}_1(r)-\tilde{F}_1(r_0)\right)\over3\sqrt{2M}}
\,,
\end{equation}%
where
\begin{equation}\label{tau011}
\tilde{F}_1(r)=\left( r-{\gamma^2\over M}\right)\sqrt{r+{\gamma^2\over 2M}}\,,
\end{equation}
and the coordinate time $t$ as a function of $r$ given by Eq. (\ref{intt}) yields
\begin{eqnarray}\label{te01}
 t\left( r\right)&=&\, \tau \left( r\right)
\pm{2\over \sqrt{2M}}\left( 2M\,(G_0(r)-G_0(r_0))\right.\\
&&\notag
\left.+A_1\,(G_1(r)-G_1(r_0))+A_2\,(G_2(r)-G_2(r_0))\right)\,, 
  \end{eqnarray}
where $r_0$ is a distance arbitrarily chosen as the starting point of the trajectory and the functions $G_j(r)$, with $j=1,2$, are given explicitly by
\begin{eqnarray}
  	\label{G012}
G_0(r)&=&\sqrt{r+{\gamma^2\over 2M}}
\,,\\
G_1(r)&=&{-1\over \sqrt{r_++{\gamma^2\over 2M}}}\tanh^{-1}\sqrt{r+{\gamma^2\over 2M}\over r_++{\gamma^2\over 2M}}
\,,\\
G_2(r)&=&{-1\over \sqrt{\rho_2+{\gamma^2\over 2M}}}\tanh^{-1}\sqrt{r+{\gamma^2\over 2M}\over \rho_2+{\gamma^2\over 2M}}\,,
  \end{eqnarray}
  
On the other hand, the motion of massive particles with $E>1$ and $L=0$ give us
\begin{equation}
    p_2(r)=(E^2-1)P_2(r),
\end{equation}
where $P_2(r):=r^2+\frac{2M}{E^2-1}r+\frac{\gamma^2}{E^2-1}=-P_0(r)$. The proper time $\tau(r)$ given by Eq. (\ref{inttau}) in terms of the radial coordinate $r$ yields

\begin{equation}\label{tau02}
\tau \left( r\right) =\pm
\left({\sqrt{p_2(r)}-\sqrt{p_2(r_0)}\over\sqrt{E^2-1} }-{M\,H(r)\over (E^2-1)^{3/2}}\right)
\,,
\end{equation}%
where
\begin{equation}\label{tau2}
H(r)=\ln\left| {\sqrt{P_2(r)}+r+{M\over E^2-1}\over \sqrt{P_2(r_0)}+r_0+{M\over E^2-1}}\right|\,,
\end{equation}
and the coordinate time $t$ as a function of $r$ given by Eq. (\ref{intt})  yields
\begin{eqnarray}\label{te02}
&&t\left( r\right)=\pm E \tau \left( r\right) \pm{E\over \sqrt{E^2-1}}
\left(2M H(r)+\right.
\,\\
\notag &&\left.+A_1(J_1(r)-J_1(r_0))+A_2(J_2(r)-J_2(r_0))\right)
\,,
  \end{eqnarray}
where the functions $J_j(r)$, with $j=1,2$, are given explicitly by
\begin{eqnarray}
  	\label{J012}
J_1(r)&=&{-1\over \sqrt{P_2(r_+)}}\cosh^{-1}{\Xi_1(r)}
\,,\\
\Xi_1(r)&=&{P_2(r_+)+(r-r_+)\left(r_++{M\over (E^2-1)}\right)\over (r-r_+)\sqrt{{M^2\over (E^2-1)^2}-{\gamma^2\over (E^2-1)}}} \,,\\
J_2(r)&=&{-1\over \sqrt{P_2(\rho_2)}}\cosh^{-1}{\Xi_2(r)}
\,,\\
\Xi_2(r)&=&{P_2(\rho_2)+(r-\rho_2)\left(\rho_2+{M\over (E^2-1)}\right)\over (r-\rho_2)\sqrt{{M^2\over (E^2-1)^2}-{\gamma^2\over (E^2-1)}}}\,,
  \end{eqnarray}

\begin{figure}[H]
	\begin{center}
\includegraphics[width=9cm]{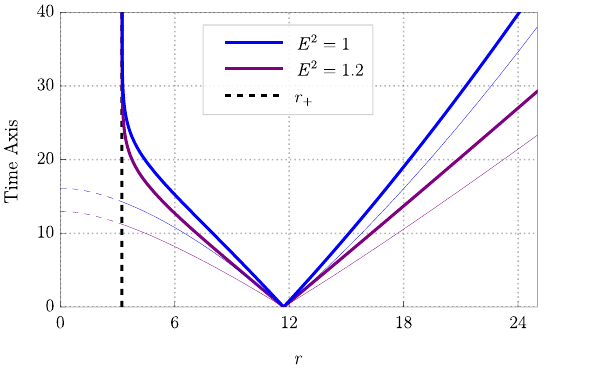}
	\end{center}
	\caption{Proper (thin curve) and coordinate (thick curve) time for the unbounded trajectories of test particles with $L=0$. We observe that the proper time that take the particles to
cross the event horizon decreases when the energy increases.
	Here, we have used the values $M=1$, $\gamma=2$, $d_0=11.708$, and $E\geq1$.}
	\label{frub}
\end{figure}

Now, in order to visualize the behavior of the proper and coordinate times, we plot their behaviors in Fig. \ref{frub}. 
We can observe that the particles
cross the event horizon in a finite proper time, but an external observer will see that the particles take an infinite (coordinate) time to do it. Moreover, the particles can escape to infinity.

\section{Final remarks}
\label{conclusion}

In this work, we considered the motion of massive neutral particles in the background of four-dimensional asymptotically flat Horndeski black holes, and we studied time-like geodesics to find and explore the geodesic structure of this spacetime. The structure of the time-like geodesics was determined analytically, and the geodesics were also studied analytically, and they were plotted to see their behavior. Our main purpose was to have a complete geodesics structure and to constrain the $\gamma$ parameter that appears in the metric function
from the solar system observations.

We first reviewed the general Horndeski theory and then, following \cite{Babichev:2017guv}, we considered an asymptotically flat Horndeski black hole which is characterized by the parameter $\gamma$, which is the ratio of the $\beta$ and $\sqrt{\eta}$ parameters, where $\beta$ indicates the strength of the interaction of the scalar field to curvature and $\eta$ is the strength of the scalar field.
To study the motion of a test massive particle in the considered background black hole, we calculated the effective potential in which the above parameters appear and also the angular momentum of the test particle.

Having the effective potential, we calculated the critical points, the scattering points, and also the inflection points to get information about the emergence of stable and unstable circular orbits, as well as the planetary orbits.  The coupling parameter $\gamma$, which indicates how strong is the interaction of matter to curvature, is a crucial parameter, giving us a lot of physics information by itself through the effective potential. We performed a detailed analysis of how the energy $E$ and the angular momentum $L$ of the test particle affect the trajectories of the particle. 
We derived a formula for the advance of the perihelia of planetary orbits that allows us to have a better accuracy between the observational value and the theoretical value of the precession of perihelion for the Solar System, giving us a constraint of the parameter $\gamma$ of Horndeski's theory.

It could be interesting to consider the motion of charged particles in order to analyze the influence of the charge of the particles on the orbits, and we hope to address it in a forthcoming work.\\

\section*{Acknowlegements}

{\bf{We thank the anonymous referee for valuable comments and suggestions.}}

\appendix
\section{Analysis of Critical, Scattering and Inflection Points of the Effective Potential}

\subsection{Critical points}
\label{CP}
First, the appearance of critical points on the potential gives us information about the emergence of stable and unstable circular orbits, as well as planetary orbits. Thus, the constraint on the first radial derivative $V_{\text{eff}}^2\,'(r)=0$, gives us the cubic polynomial
\begin{eqnarray} \label{polyrc}
    r^3+a_cr^2+b_cr+c_c=0\,,
\end{eqnarray}
where
\begin{equation}
    a_c= \frac{\gamma^2 - L^2}{M}\,, \quad 
    b_c=3L^2\,,\quad 
    c_c= \frac{2L^2\gamma^2}{M}\,,
\end{equation}
the index $c$ denotes the {\it critical points}. Through the Tschirnhaus transformation $r=x-\frac{a_c}{3}$ and the cubic sine equation, we find the roots
\begin{eqnarray}
    r_{c,1}&=& 2 \sqrt{\frac{p_c}{3}} \sin \left( \frac{\theta_c}{3} \right) - \frac{a_c}{3} ,\\
    r_{c,2}&=& 2 \sqrt{\frac{p_c}{3}} \sin \left( \frac{\theta_c}{3} + \frac{2\pi}{3}\right) - \frac{a_c}{3}  ,\\
    r_{c,3}&=& 2 \sqrt{\frac{p_c}{3}} \sin \left( \frac{\theta_c}{3} + \frac{4\pi}{3}\right) - \frac{a_c}{3},
\end{eqnarray}
where the angle $\theta_c$ is
\begin{eqnarray}
    \theta_c &=& \arcsin\left( \frac{q_c}{2} \sqrt{ \left( \frac{3}{p_c} \right)^3} \right)\,,
\end{eqnarray}
and the coefficients 
\begin{equation}
    p_c =  \frac{a_c^2}{3} - b_c\,, \quad
    q_c = c_c + \frac{2a_c^3}{3^3} - \frac{a_c b_c }{3}\,,
\end{equation}
where $r_{c,1} =: r_U$ and $r_{c,2} =: r_{S}$ for unstable and stable circular radii, respectively, and $r_{c,3}$ is a negative root without physical interest. It is important to determine the range of angular momentum for which these types of orbits can occur. For that, it is enough to study the discriminant of the cubic polynomial (\ref{polyrc}), whose expression is equal to
\begin{eqnarray} \label{disrc}
    \triangle_{c}(L) = \alpha_{c} L^2 P_c(L)\,,
\end{eqnarray}
where 
\begin{eqnarray} \label{polyamc}
    P_c(L) &=& L^6 + \bar{a}_{c} L^4 + \bar{b}_{c} L^2+\bar{c}_{c}\,,
\end{eqnarray}
and the coefficients are 
\begin{eqnarray}
    \alpha_{c} &=& \frac{9 M^2 + 8 \gamma^2}{   M^4}\,, \\
    \bar{a}_c &=& -6  \frac{(18M^4+21 M^2 \gamma^2 + 4 \gamma^4)}{(9M^2+8\gamma^2)}\,, \\
    \bar{b}_c &=& 3  \gamma^4 
 \frac{(3M^2+8\gamma^2)}{(9M^2+8\gamma^2)}\,, \\
    \bar{c}_c &=& -\frac{8  \gamma^8}{9 M^2 + 8 \gamma^2}\,.
\end{eqnarray}
We can observe that (\ref{polyamc}) is a polynomial of degree six, which becomes a cubic polynomial after the change of variable $K = L^2$. The discriminant of this cubic polynomial is negative, which means that it has only one real root, while the other two roots of $K$ are complex and conjugate to each other, since the coefficients of the polynomial are real. Furthermore, by applying Descartes' rule of signs to the original degree-six polynomial, we conclude that for the angular momentum $L$ there are only two real roots with opposite signs and four complex roots, two of which are conjugates of each other. This positive real root has the form
\begin{eqnarray}
    L_C =\sqrt{ 2 \sqrt{\frac{\bar{p}_{c}}{3}} \sin \left( \frac{\bar{\theta}_c}{3} +  \frac{2\pi}{3}\right) - \frac{\bar{a}_c}{3}}\,,
\end{eqnarray}
where the angle has the form
\begin{eqnarray}
    \bar{\theta}_{c}= \arcsin\left( \frac{\bar{q}_c}{2} \sqrt{ \left( \frac{3}{\bar{p}_c} \right)^3} \right)\,,
\end{eqnarray}
and the coefficients are
\begin{equation}
    \bar{p}_c = \frac{\bar{a}_c^2}{3} - \bar{b}_c\,, \quad
    \bar{q}_c = \bar{c}_c + \frac{2\bar{a}_c^3}{3^3} - \frac{\bar{a}_c \bar{b}_c }{3}\,.
\end{equation}

Hence, we can conclude that
in the range $ L \in [0, L_C) $, there are no circular orbits, since we have two complex radial roots from the negativity of the discriminant (\ref{disrc}) and one negative real radial root by the Descartes rule of signs on the polynomial equation (\ref{polyrc}), this last negative root remains in the other two next cases, so we will ignore it. When $ L = L_C $, the discriminant is zero, so we obtain two equal positive real roots $r_U=r_S=r_{\text{ISCO}}$ that correspond to the innermost stable circular orbit. 
In the range $ L \in (L_C, \infty) $, the discriminant is positive, so there are two different positive real roots, then there are two circular orbits (unstable and stable). Furthermore, by parity symmetry on the polynomial (\ref{polyamc}), when $ L $ is negative, the behavior is analogous by considering the negative real root $ -L_C $.

\subsection{Scattering points}
\label{SP}
It is also possible to identify a range of values of $L$ for which particles undergo scattering. To determine this range, it is sufficient to analyze the values of $L$ for which the effective potential reaches or exceeds its asymptotic value
\begin{eqnarray}
    V_{\text{eff}}^{\infty} := \lim\limits_{r \to +\infty} V_{\text{eff}} (r) = 1,
\end{eqnarray}
namely, we must study the polynomial equation
\begin{eqnarray} \label{polyrs}
    r^3 + a_{s} r^2 + b_{s} r + c_{s} \leq 0\,,
\end{eqnarray}
where
\begin{equation}
    a_{s} = \frac{\gamma^2  - L^2}{2  M }\,, \quad
    b_{s} =L^2\,, \quad
    c_{s} = \frac{L^2 \gamma^2}{2M}\,.
\end{equation}
Here, the index $s$ denotes the {\it scattering curves}. Similarly to the critical points analysis, the roots of this equation are
\begin{eqnarray}
    r_{s,1}&=& 2 \sqrt{\frac{p_s}{3}} \sin \left( \frac{\theta_s}{3} \right) - \frac{a_s}{3}\,,\\
    r_{s,2}&=& 2 \sqrt{\frac{p_s}{3}} \sin \left( \frac{\theta_s}{3} + \frac{2\pi}{3}\right) - \frac{a_s}{3}\,,\\
    r_{s,3}&=& 2 \sqrt{\frac{p_s}{3}} \sin \left( \frac{\theta_s}{3} + \frac{4\pi}{3}\right) - \frac{a_s}{3}\,,
\end{eqnarray}
where

\begin{equation}
    \theta_s = \arcsin\left( \frac{q_s}{2} \sqrt{ \left( \frac{3}{p_s} \right)^3} \right)\,,
\end{equation}
\begin{equation}
p_s =  \frac{a_s^2}{3} - b_s\,, \quad
    q_s = c_s + \frac{2a_s^3}{3^3} - \frac{a_s b_s }{3}\,.
\end{equation}
In the same way, through the discriminant
\begin{eqnarray} \label{disrS}
    \triangle_{s}(L) = \alpha_{s} L^2 P_s(L)\,,
\end{eqnarray}
where 
\begin{eqnarray} \label{polyamS}
    P_s(L) &=& L^6 + \bar{a}_s L^4 + \bar{b}_s L^2 + \bar{c}_s\,,
\end{eqnarray}
and the coefficients are
\begin{eqnarray}
    \alpha_{s} &=& \frac{M^2 + \gamma^2}{ 4 M^4}\,, \\
    \bar{a}_s &=& - \frac{(16M^4+20 M^2 \gamma^2 + 3 \gamma^4)}{(M^2+\gamma^2)}\,, \\
    \bar{b}_s &=&  - \gamma^4 
    \frac{(8M^2-3\gamma^2)}{(M^2+\gamma^2)}, \\
    \bar{c}_s &=& -\frac{ \gamma^8}{( M^2 +  \gamma^2)}\,,
\end{eqnarray}
we obtain the angular momentum so that the particle has the possibility of having scattering orbits with
\begin{eqnarray}
    L_S =\sqrt{ 2 \sqrt{\frac{\bar{p}_s}{3}} \sin \left( \frac{\bar{\theta}_s}{3} +  \frac{2\pi}{3}\right) - \frac{\bar{a}_s}{3}}\,,
\end{eqnarray}
with
\begin{equation}
    \bar{\theta}_{s} = \arcsin\left( \frac{\bar{q}_s}{2} \sqrt{ \left( \frac{3}{\bar{p}_s} \right)^3} \right)\,,
\end{equation}
\begin{equation}
    \bar{p}_s = \frac{\bar{a}_s^2}{3} - \bar{b}_s\,,\quad 
    \bar{q}_s = \bar{c}_s + \frac{2\bar{a}_s^3}{3^3} - \frac{\bar{a}_s \bar{b}_s }{3}\,.
\end{equation}
Following the same analysis as for the critical points, first, by Descartes' rule of signs, there is one negative real radial root for (\ref{polyrs}), for every angular momentum value. On the other hand, in the range $ L \in [0, L_S) $, there are no scattering orbits, since the discriminant of the scattering polynomial (\ref{polyrs}) is negative, leaving us with two complex radial roots. When $ L = L_S $, the discriminant is zero, so we obtain two equal positive real radial roots, leading to only one scattering region at the energy $ V_{\text{eff}}^{\infty} $, located at the same position as the unstable circular orbit. In the range $ L \in (L_S, \infty) $, the discriminant is positive, so there are two different positive real radial roots. Hence, there is a scattering zone between the energy $ V_{\text{eff}}^{\infty} $ and the energy of the unstable circular orbit. For negative angular momentum values, the analysis is analogous.

\subsection{Inflection points}
\label{IP}
The inflection point of the effective potential also gives information about the orbits, since from (\ref{g9}) we have the relation between the particle radial acceleration and the second order derivative of the effective potential
\begin{equation} \label{acel}
    a_r = \frac{d^2r}{d\tau^2} = - \frac{1}{2} V_{\text{eff}}^{2}\, '(r)\,.
\end{equation}
In this way, the study of the constraint $V_{\text{eff}}^{2} \,''(r)=0$ gives information about the radial distance and the angular momentum where the acceleration reaches its highest and lowest value. Then, from $V_{\text{eff}}^{2} \,''(r)=0$, we obtain the cubic polynomial
\begin{eqnarray}  \label{polyri}
    r^3+a_{I}r^2+b_{I} r+c_{I}=0\,,
\end{eqnarray}
where
\begin{equation}
    a_{I} = 3\frac{\left( \gamma ^2 - L^2\right)}{2  M}\,, \quad
    b_{I} = 6 L^2\,, \quad
    c_{I} = \frac{5 \gamma ^2 L^2}{ M}\,.
\end{equation}
Here, the index $I$ {\bf{denotes}} the {\it  inflection points}. Hence, the roots of this equation are
\begin{eqnarray}
    r_{I,1}&=& 2 \sqrt{\frac{p_I}{3}} \sin \left( \frac{\theta_I}{3} \right) - \frac{a_I}{3}\,,\\
    r_{I,2}&=& 2 \sqrt{\frac{p_I}{3}} \sin \left( \frac{\theta_I}{3} + \frac{2\pi}{3}\right) - \frac{a_I}{3}\,,\\
    r_{I,3}&=& 2 \sqrt{\frac{p_I}{3}} \sin \left( \frac{\theta_I}{3} + \frac{4\pi}{3}\right) - \frac{a_I}{3}\,,
\end{eqnarray}
where
\begin{equation}
    \theta_I = \arcsin\left( \frac{q_I}{2} \sqrt{ \left( \frac{3}{p_I} \right)^3} \right)\,, 
\end{equation}
\begin{equation}
    p_I =  \frac{a_I^2}{3} - b_I\,, \quad
    q_I = c_I + \frac{2a_I^3}{3^3} - \frac{a_I b_I }{3}\,.
\end{equation}
The discriminant of this cubic polynomial (\ref{polyri}) is
\begin{eqnarray} \label{disrI}
    \triangle_{I}(L) = \alpha_{I} L^2 P_I(L)\,,
\end{eqnarray}
where 
\begin{eqnarray} \label{polyamI}
    P_I(L) &=& L^6 + \bar{a}_I L^4 + \bar{b}_I L^2+\bar{c}_I\,,
\end{eqnarray}
and the coefficients
\begin{eqnarray}
    \alpha_{I} &=& \frac{27 \left(5 \gamma ^2+6 M^2\right)}{2 M^4}\,, \\
    \bar{a}_I &=& -  \frac{\left(15 \gamma ^4+64 M^4+72 \gamma ^2 M^2\right)}{5 \gamma ^2+6 M^2}\,, \\
    \bar{b}_I &=&   \gamma ^4\frac{  \left(15 \gamma ^2+16 M^2\right)}{5 \gamma ^2+6 M^2}\,, \\
    \bar{c}_I &=& -\frac{5 \gamma ^8  }{5 \gamma ^2+6 M^2}\,.
\end{eqnarray}
Thus, we obtain the angular momentum value for which the particle can experience both a maximum and a minimum acceleration. It takes the form
\begin{eqnarray}
    L_I =\sqrt{ 2 \sqrt{\frac{\bar{p}_I}{3}} \sin \left( \frac{\bar{\theta}_I}{3} +  \frac{2\pi}{3}\right) - \frac{\bar{a}_I}{3}},
\end{eqnarray}
where
\begin{equation}
    \bar{\theta}_{I} = \arcsin\left( \frac{\bar{q}_I}{2} \sqrt{ \left( \frac{3}{\bar{p}_I} \right)^3} \right)\,, 
\end{equation}
\begin{equation}
    \bar{p}_I = \frac{\bar{a}_I^2}{3} - \bar{b}_I\,,\quad
    \bar{q}_I = \bar{c}_I + \frac{2\bar{a}_I^3}{3^3} - \frac{\bar{a}_I \bar{b}_I }{3}.
\end{equation}
As in the previous two analyses, one of the radial roots remains a negative real root for all values of the angular momentum, while the behavior of the other two is governed by the following conditions: if $ L \in [0, L_I) $, there are no radial zones where the radial acceleration becomes zero, since the discriminant (\ref{disrI}) is negative, leaving us with two complex roots for the inflection points. For $ L = L_I $, there is one radial zone where the radial acceleration becomes zero, which occurs because the discriminant is zero, that is, there are two equal positive real roots. In the range $ L \in (L_I, \infty) $, the discriminant is positive, so the radial acceleration is zero in two different places.

\subsection{Analysis}

Combining the three analyses above, we find that $ L_I \leq L_C $, since the discriminants $ \triangle_c(L) $ and $ \triangle_I(L) $ imply that the polynomials in Eqs.~(\ref{polyamc}) and (\ref{polyamI}) satisfy the inequality $ P_c(L) \leq P_I(L) $. This holds because the difference $ P_c(L) - P_I(L) $ is explicitly negative. Therefore, evaluating the polynomial inequality at $ L = L_C $ yields the corresponding inequality for the angular momentum. Similarly, it follows that $ L_C \leq L_S $. In general, for positive angular momentum, the inequality $ L_I \leq L_C \leq L_S $ holds, while for negative angular momentum, the relation $ -L_S \leq -L_C \leq -L_I $ is satisfied.

\begin{figure}[H]
    \begin{center}
        \includegraphics[width=9cm]{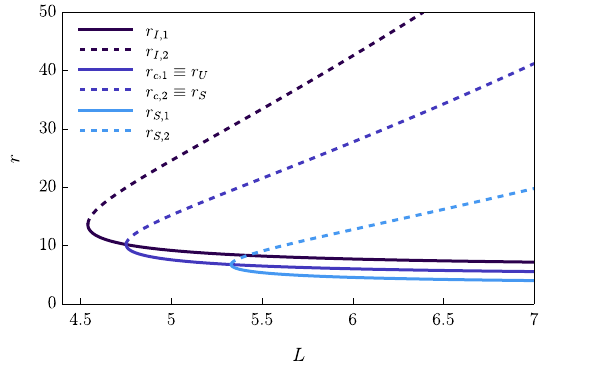}
    \end{center}
    \caption{
    Three distinct curves, each in a different color, represent the behavior of a massive particle for angular momentum values equal to or greater than $L_{I} = 4.541$, $L_{C}= 4.751$, and $L_{S} = 5.330$, respectively, with $M = 1$ and $\gamma = 2$. Each curve corresponds to the emergence of two real roots in the radial equation. The solid lines denote the decreasing root as a function of $L$, while the dashed lines denote the increasing root.}
    \label{frootsL}
\end{figure}

Furthermore, as the angular momentum increases and reaches the value $ L_I $, two physical radial regions emerge from the three roots of the inflection point polynomial (Eq.~\eqref{polyri}). This behavior is illustrated in Fig.~\ref{frootsL}, where two branches can be observed: in one branch (dashed curve), the radius of the inflection point increases with increasing $ L $, while in the other (solid curve), the radius decreases as $ L $ increases.
A similar behavior is seen for $ L_C $, associated with the critical point polynomial (Eq.~\eqref{polyrc}). From $ r_{\text{ISCO}} $, the radius of the stable circular orbit increases (dashed curve) with increasing $ L $, whereas the radius of the unstable circular orbit decreases (solid curve).
The same qualitative behavior also occurs for $ L_S $, as obtained from the scattering point polynomial (Eq.~\eqref{polyrs}).

\begin{figure}[H]
    \begin{center}
        \includegraphics[width=9cm]{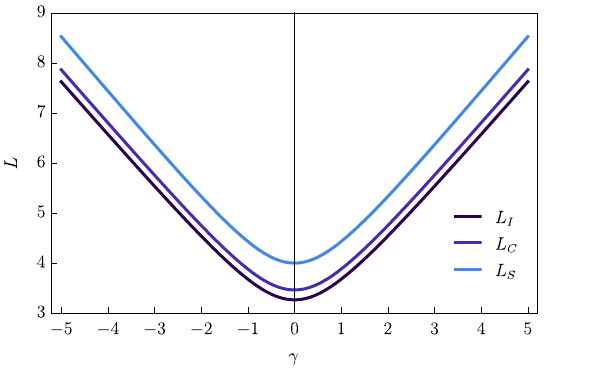}
    \end{center}
    \caption{Behavior of $L_S$, $L_C$, and $L_I$ as a function of the coupling parameter $\gamma$ for a massive particle and fixed $M=1$.}
    \label{fangularmomentum}
\end{figure}

In Fig.~\ref{fangularmomentum}, the three angular momentum values $L_I$, $L_C$, and $L_S$ increase as $\gamma > 0$ increases. This indicates that a massive particle requires greater angular momentum to access different types of orbits, such as unstable circular, stable circular, planetary, or scattering orbits.

\end{document}